\documentclass[a4paper,12pt, twoside]{article}
\usepackage[english]{babel}
\usepackage{rotating}
\usepackage{epsfig}
\usepackage[centertags]{amsmath}
\usepackage{amssymb,amscd,./styles/amsatdef}
\usepackage{array}
\usepackage{multirow}
\usepackage{supertabular}
\usepackage{dcolumn}
\usepackage{xspace}
\def\zero{{\scriptscriptstyle 0}}

\def\Z0{\ensuremath{Z^\zero}}

\def\SU2U1{{\rm SU}(2)\times{\rm U}(1)}

\def\exp{{\rm exp}}

\mathchardef\qsm=63
\mathchardef\pls=43
\mathchardef\mns=512
\mathchardef\plm=518
\mathchardef\eql=61
\mathchardef\smallleft=300
\mathchardef\smallright=301
\mathchardef\perslsh=47
\mathchardef\les=316
\mathchardef\gre=318
\mathchardef\leq=532
\mathchardef\grq=533
\chardef\usc=95
\chardef\til=126

\def\sqr#1#2#3{{\vcenter{\hrule height.#3ex\hbox{\vrule width.#2ex height#1ex
    \kern#1ex\vrule width.#3ex}\hrule height.#2ex}}}

\def\angleto{\vrule width.035em height2.1ex depth-.56ex\unskip\kern-.6ex\to}
\def\perchc#1{{\raise.4ex\hbox{$\mkern4mu#1{\it\perslsh}_
             {\mkern-5mu\scriptscriptstyle{{\rm o}\!{\rm o}}}^
             {\mkern-12.8mu\scriptscriptstyle{\rm o}}$}}}

\catcode`\@=11 
\def\parenbar{\mathpalette\p@renb@r}
\def\p@renb@r#1#2{\vbox{%
  \ifx#1\scriptscriptstyle \dimen@.7em\dimen@ii.2em\else
  \ifx#1\scriptstyle \dimen@.8em\dimen@ii.25em\else
  \dimen@1em\dimen@ii.4em\fi\fi \offinterlineskip
  \ialign{\hfill##\hfill\cr
    \vbox{\hrule width\dimen@ii}\cr
    \noalign{\vskip-.3ex}%
    \hbox to\dimen@{$\mathchar300\hfil\mathchar301$}\cr
    \noalign{\vskip-.3ex}%
    $#1#2$\cr}}}
\catcode`\@=12 

\newbox\struttbox
\setbox\struttbox=\hbox{\vrule height1.65ex depth.485ex width0pt}
\def\strutt{\relax\ifmmode\copy\struttbox\else\unhcopy\struttbox\fi}
\def\stru#1#2{\relax\ifmmode\hbox{\vrule height#1 depth#2 width0pt}
\else\vrule height#1 depth#2 width0pt\fi}

\def\ronum#1{\uppercase\expandafter{\romannumeral#1}}
\def\ronuml#1{\expandafter{\romannumeral#1}}

\DeclareMathAlphabet{\mathbf}{OT1}{cmr}{bx}{sl}

\def\P{{\rm I\kern-.15em P}}
\def\xp{x_{\P}}
\def\q2{Q^{2}}

\newcommand{\sgeq} {\raisebox{-.6ex}{${\textstyle\stackrel{>}{\sim}}$}}
\newcommand {\avT} {$\langle T \rangle$}
\newcommand {\avS} {$\langle S \rangle$}
\newcommand {\avM} {$\langle M_X \rangle$}
\newcommand {\avthtt} {$\langle \theta_{\rm Thrust} \rangle$}
\newcommand {\avptt} {$\langle p_{\rm T}^{\rm Thrust} \rangle$}

\newcommand{\qbarq}{\mbox{$q\bar{q}$}}                
\newcommand{\qbarqg}{\mbox{$q\bar{q}g$}}              

\newcommand{\ZcoosysA}{%
The ZEUS coordinate system is a right-handed Cartesian system, with the $Z$
axis pointing in the proton beam direction, referred to as the ``forward
direction'', and the $X$ axis pointing left towards the center of HERA.
The coordinate origin is at the nominal interaction point.\xspace}
\newcommand{\Zpsrap}{%
The pseudorapidity is defined as $\eta=-\ln\left(\tan\frac{\theta}{2}\right)$,
where the polar angle, $\theta$, is measured with respect to the proton beam
direction.\xspace}

 {\end{list}}
 {\end{list}}
 {\end{list}}

\catcode`\@=11 
\newlength{\@fninsert}
\newlength{\@fnwidth}
\renewcommand{\@makefntext}[1]%
  {\noindent\makebox[\@fninsert][r]{\@makefnmark}\hfil%
  \parbox[t]{\@fnwidth}{#1}}
\catcode`\@=12 
\newlength{\localtextwidth}
\catcode`\@=11 
\newsavebox{\tmpbox}
\newlength{\@captionmargin}
\newlength{\@captionwidth}
\newlength{\@captionitemtextsep}
\renewcommand{\@makecaption}[2]%
  {%
   \vspace{10.pt}
   \setlength{\@captionwidth}{\localtextwidth}
   \addtolength{\@captionwidth}{-\@captionmargin}
   \sbox{\tmpbox}{{\bf #1:}{\it #2}}%
   \ifthenelse{\lengthtest{\wd\tmpbox > \@captionwidth}}%
   {\centerline{\parbox[t]{\@captionwidth}%
   {\tolerance=2000\normalsize%
    {\bf #1:}\hspace{\@captionitemtextsep}{\it #2}}}}%
   {\centerline{{\bf #1:}\kern1.em{\it #2}}}}
\catcode`\@=12 
\catcode`\@=11 
\renewcommand\section{\@startsection{section}{1}{\z@}%
                                   {-3.5ex \@plus -1ex \@minus -.2ex}%
                                   {2.3ex \@plus.2ex}%
                                   {\normalfont\Large\bfseries}}
\renewcommand\subsection{\@startsection{subsection}{2}{\z@}%
                                   {-3.25ex\@plus -1ex \@minus -.2ex}%
                                   {1.5ex \@plus .2ex}%
                                   {\normalfont\large\bfseries}}
\renewcommand\subsubsection{\@startsection{subsubsection}{3}{\z@}%
                                   {-3.25ex\@plus -1ex \@minus -.2ex}%
                                   {1.5ex \@plus .2ex}%
                                   {\normalfont\large\bfseries}}
\renewcommand\paragraph{\@startsection{paragraph}{4}{\z@}%
                                   {3.25ex \@plus1ex \@minus.2ex}%
                                   {1.2ex \@plus .2ex}%
                                   {\normalfont\normalsize\bfseries}}
\catcode`\@=12 

\newsavebox{\sesbox}
\newlength{\seslen}

\begin{document}
\selectlanguage{english}

\thispagestyle{empty}
\title{
{}
{}
{}
{}
{}
\vskip2.2cm
\bf\LARGE Properties of hadronic final states \\
\bf\LARGE in diffractive deep inelastic $ep$ scattering at HERA
}                                                       
                    
\author{ZEUS Collaboration}
\date{}
\maketitle

\vspace{-10.5cm}
\begin{flushleft}
\tt DESY 01-097 \\
July 2001 \\
\end{flushleft}
\vspace{+10cm}

\vfill
\centerline{\bf Abstract}
\vskip4.mm
\centerline{
  \begin{minipage}{15.cm}
        \noindent
        Characteristics of the hadronic final state of diffractive deep inelastic scattering
        events, $ep \rightarrow eXp$,  were studied in the kinematic range 
	$4 < M_X < 35$ GeV, $4 < Q^2 < 150$ GeV$^2$, 
	$70 < W < 250$ GeV and $0.0003 < \xp < 0.03$ with the ZEUS detector at 
	HERA using an integrated luminosity of 13.8 pb$^{-1}$.
        The events were tagged by identifying the diffractively scattered
        proton using the leading proton spectrometer.
        The properties of the hadronic final state, $X$, were studied 
	in its center-of-mass frame
	using thrust, thrust angle,        sphericity, 
        energy flow, transverse energy flow and ``seagull'' distributions.
	As the invariant mass of the 
	system increases, the final state becomes more collimated, more aligned  
	and more asymmetric in the average transverse momentum with respect
	to the direction of the virtual photon.
	Comparisons of the properties of the hadronic final state with predictions 
	from various Monte Carlo model generators suggest that the final state is 
	dominated by $q\bar{q}g$ states at the parton level.
        \newline
  \end{minipage}
  }

\vfill

\thispagestyle{empty}
\pagestyle{plain}
\newpage
%
%
%
%
\topmargin-1.cm
\evensidemargin-0.3cm
\oddsidemargin-0.3cm                                                                              
\textheight 680pt
\parindent0.cm
\parskip0.3cm plus0.05cm minus0.05cm
\def\3{\ss}                                                                                        
\newcommand{\address}{ }                                                                           
\pagenumbering{Roman}                                                                              
                                                   %
\begin{center}                                                                                     
{                      \Large  The ZEUS Collaboration              }                               
\end{center}                                                                                       
  S.~Chekanov,                                                                                     
  M.~Derrick,                                                                                      
  D.~Krakauer,                                                                                     
  S.~Magill,                                                                                       
  B.~Musgrave,                                                                                     
  A.~Pellegrino,                                                                                   
  J.~Repond,                                                                                       
  R.~Yoshida\\                                                                                     
 {\it Argonne National Laboratory, Argonne, Illinois 60439-4815}                            
\par \filbreak                                                                                     
  M.C.K.~Mattingly \\                                                                              
 {\it Andrews University, Berrien Springs, Michigan 49104-0380}                                    
\par \filbreak                                                                                     
  P.~Antonioli,                                                                                    
  G.~Bari,                                                                                         
  M.~Basile,                                                                                       
  L.~Bellagamba,                                                                                   
  D.~Boscherini$^{   1}$,                                                                          
  A.~Bruni,                                                                                        
  G.~Bruni,                                                                                        
  G.~Cara~Romeo,                                                                                   
  L.~Cifarelli$^{   2}$,                                                                           
  F.~Cindolo,                                                                                      
  A.~Contin,                                                                                       
  M.~Corradi,                                                                                      
  S.~De~Pasquale,                                                                                  
  P.~Giusti,                                                                                       
  G.~Iacobucci,                                                                                    
  G.~Levi,                                                                                         
  A.~Margotti,                                                                                     
  T.~Massam,                                                                                       
  R.~Nania,                                                                                        
  F.~Palmonari,                                                                                    
  A.~Pesci,                                                                                        
  G.~Sartorelli,                                                                                   
  A.~Zichichi  \\                                                                                  
  {\it University and INFN Bologna, Bologna, Italy}
% ...
\newpage                                                                                           
$^{\    1}$ now visiting scientist at DESY \\                                                      
$^{\    2}$ now at Univ. of Salerno and INFN Napoli, Italy \\                                      
% ...
                                                           %
                                                           %
                                                           %

\newpage

\topmargin-1.5cm
\evensidemargin-0.3cm
\oddsidemargin-0.3cm
\textwidth 16.cm
\textheight 650pt
\parindent0.cm
\parskip0.3cm plus0.05cm minus0.05cm

\pagenumbering{arabic} 
\setcounter{page}{1}


\section{Introduction}

   A class of deep inelastic scattering (DIS)
events has been observed at HERA that have the characteristics of diffractive
interactions.  
These events have 
 a large rapidity gap~\cite{h1zeusrapgap} between the recoil-proton system and the
produced hadronic system, and a small momentum transfer 
to the proton~\cite{lps_f2}.
The events can be pictured
in terms of the $t$-channel exchange
of an object that carries the quantum numbers of the vacuum, called 
the Pomeron ($\P$), see Fig.~\ref{fg:feynman}(a). However,
the nature of the Pomeron in DIS is at present far from
clear.
Measurements by the H1~\cite{h1f2d4, h1recentdiff, h1hadron} and ZEUS~\cite{zeusrecentdiff}
 collaborations have shown that, in QCD-inspired
 models of the diffractive process, the Pomeron can be described as an 
object whose partonic composition is dominated by gluons. 
Alternatively, the diffractive process can be described by the dissociation of the
virtual photon into a $q\bar{q}$ or $q\bar{q}g$ final state that interacts with the 
proton by the exchange of a gluon ladder~\cite{QCDmodels}. 

The study of the hadronic final state in $e^+e^-$ annihilation  \cite{eedata}
has been a powerful tool in gaining information about the underlying partonic state. 
Similarly, the study of the partonic content of the hadronic final state
in diffraction is a natural way to explore the dynamics of diffraction.

In this paper, a study is reported of the hadronic system, $X$, produced in 
the DIS process $ep \rightarrow eXp$, where the diffractively scattered
proton stays intact. 
The proton was detected and its three-momentum measured in the 
leading proton spectrometer (LPS)~\cite{lps}.
Diffractive events are defined, for the purpose of this paper, as those
events which contain a proton with more than $97 \%$ of the initial 
proton beam energy.
Previous results on hadronic final states in diffractive events at HERA have been
obtained with the requirement of a large rapidity gap between the observed
hadronic system and the scattered proton~\cite{zeuseventshape, h1eventshape}.
The results obtained with rapidity-gap events were either defined in a reduced phase space
by imposing a cut in rapidity~\cite{zeuseventshape, pomprobe},
or Monte Carlo simulated events were used to extrapolate the characteristics of diffractive
events over the areas of phase space
removed by the rapidity-gap cut \cite{h1eventshape}.
By using the scattered proton to tag diffractive events, there is no need to
rely on Monte Carlo generators
to model correctly the part of the final state removed by the rapidity cuts,
and the full angular coverage of the central detector can be used.

The properties of the hadronic system, $X$, were studied in terms of global 
event-shape variables such as
thrust and sphericity in the center-of-mass (CMS) frame of $X$.
This is analogous to the studies of global event-shape variables in
$e^+e^-$ annihilation~\cite{eedata}
as a function of the CMS energy 
and to the analysis
that led to the interpretation of three-jet events in terms of
gluon bremsstrahlung \cite{gluondiscovery}. 
In addition to global event-shape variables, the properties of the diffractive events 
were described in terms of inclusive distributions such as energy flow.

%
%
\section{Experimental setup}
\label{sec:exp}


The data were recorded in 1997 with the ZEUS detector
and correspond to an integrated luminosity of $13.8 \pm 0.3$~pb$^{-1}$. 
A detailed description of the ZEUS detector can be found
elsewhere~\cite{zeus}. A brief outline of the main detector components 
most relevant for this analysis is given below. 
Charged particles are tracked by the central tracking detector (CTD)~\cite{ctd},
which operates in a magnetic field of 1.43~T provided by a thin
superconducting coil. The CTD consists of 72 cylindrical drift chamber
layers, organized in 9 superlayers covering the polar 
angle\footnote{\ZcoosysA \Zpsrap} region \mbox{$15^\circ < \theta < 164^\circ$.}
The transverse-momentum 
resolution for 
full-length tracks
is $\sigma(p_T)/p_T=0.0058p_T\oplus 0.0065 \oplus 0.0014/p_T$,
with $p_T$ in GeV.
The high-resolution uranium-scintillator calorimeter (CAL) \cite{cal}
consists of three parts: the forward (FCAL), the barrel (BCAL) and the
rear (RCAL) calorimeters. Each part is subdivided transversely into
towers and longitudinally into one electromagnetic section (EMC) and
either one (in RCAL) or two (in BCAL and FCAL) hadronic sections (HAC).
The smallest subdivision of the calorimeter is called a cell.
The CAL energy resolutions, as measured under
test beam conditions, are $\sigma(E)/E=0.18/\sqrt{E}$ for electrons and
$\sigma(E)/E=0.35/\sqrt{E}$ for hadrons ($E$ in GeV).

The LPS~\cite{lps} detects
charged particles scattered at small angles and carrying a substantial fraction
of the incoming proton momentum; these particles remain in the beampipe
and their trajectory is measured by a system of position-sensitive silicon 
micro-strip detectors very close to the proton beam. 
The track deflection induced by the magnets in the 
proton beamline is used for the momentum analysis of the scattered proton.
The LPS consists of six detector stations, 
S1 to S6, placed along the beamline in the direction of the outgoing 
protons, at $Z=23.8$~m, 40.3~m, 44.5~m, 
63.0~m, 81.2~m and 90.0~m from the interaction point, respectively.
In this analysis, only the stations S4 to S6 were used. These stations
consist of two halves, each equipped with an assembly of 
six parallel planes of silicon micro-strip detectors 
which can be inserted to a position near the proton
beam. Each detector plane has an elliptical cutout which 
follows the profile of the 
10$\sigma$ envelope of the beam, where~$\sigma$ is the standard deviation 
of the spatial distribution of the beam in the transverse plane.
The acceptance of stations S4 to S6 of the LPS for protons close to the beam
energy is a few per cent and approximately uniform for
$0.075 < |t| < 0.35$ GeV$^2$.
The LPS longitudinal momentum resolution is $\sigma(p_Z)/p_Z = 5\times 10^{-6} p_Z$ ($p_Z$ in GeV). 
The transverse momentum resolutions, dominated by the proton beam emittance,
are $\sigma_{p_X} = 35$ MeV and $\sigma_{p_Y} = 90$ MeV.

%
%

\section{Kinematic variables and event reconstruction} 
\label{sec:rec}

\subsection{Kinematic variables}

The event kinematics  of  DIS  processes can be described
by the negative squared four-momentum transfer
at the lepton vertex,   
$Q^2=-q^2=-(k-k^{'})^2$ ($k$ and $k^{'}$ denote the four-momenta of the
initial- and final-state positrons, respectively), 
and the Bjorken
scaling variable, $x=Q^2/(2 P \cdot  q)$, 
where $P$ is the
four-momentum of the proton.
The fraction of the energy transferred   
to the proton in its rest frame, $y$,    
is related to these two variables
by  $y = P \cdot q/ (P \cdot k) \simeq Q^2/xs$,
where $\sqrt{s}$ is the positron-proton
CMS energy. The CMS energy of the $\gamma^*p$ system, $W$, is
given by $W^2 = (q + P)^2 = Q^2(1-x)/x + M_p^2$, where $M_p$ denotes the proton mass.
Unless otherwise noted, the double angle method~\cite{damethod} is used
to reconstruct the kinematic variables from the measured quantities.

Two more kinematic variables are needed to describe 
a diffractive DIS event, $e(k) + p(P)\to e'(k') + X + p'(P')$,
where the scattered proton, $p'$, has four-momentum $P^\prime = (E_{p'}, P'_X, P'_Y, P'_Z)$.
The squared four-momentum transfer at the proton vertex is given by
\begin{eqnarray}
t = (P-P')^2 \simeq -P_T^2,
\nonumber
\end{eqnarray}
where $P_T^2 = P_X^{\prime 2} + P_Y^{\prime 2}$,
and the fraction of the beam momentum retained by the final proton is
\begin{eqnarray}
x_L = \frac{|\bf{P'}|}{|\bf{P}|}\simeq \frac{E_{p'}}{E_p},
\nonumber
\end{eqnarray}
where $E_p = 820$~GeV denotes the incident proton beam energy. 
Both $t$ and $x_L$ are measured with the LPS. 
Other useful variables
in diffractive DIS are:
\begin{eqnarray}
\xp = \frac{(P-P')\cdot q}{P\cdot q} = \frac{M_X^2 + Q^2 - t}{W^2 + Q^2 - 
M_p^2} \simeq 1 - x_L
\nonumber
\end{eqnarray}
and
\begin{eqnarray}
\beta = \frac{Q^2}{2(P-P')\cdot q} = \frac{x}{\xp} \simeq \frac{Q^2}{ 
M_X^2 + Q^2},
\nonumber
\label{eq:beta}
\end{eqnarray}
where $M_X$ is the invariant mass of the final-state, $X$, determined as described below.
In the Pomeron-exchange picture, $\xp$ is the fraction of the 
proton's momentum carried by the Pomeron.  For a Pomeron with partonic 
constituents, $\beta$  is then the fraction of the Pomeron's
momentum carried by the parton that absorbs the virtual photon.

\subsection{Reconstruction of $M_X$}

Two methods were used in this analysis to determine the invariant mass of the system $X$.
The first method measured the mass directly from $X$. 
The individual objects
belonging to $X$ were reconstructed
by combining charged tracks
measured in the CTD and neutral energy clusters measured in the calorimeter
into energy flow objects (EFOs~\cite{zeusrecentdiff}).
In this procedure, the tracks and clusters associated with the scattered positron
were not used.
The four-vectors of the EFOs were calculated assuming
the pion mass. The invariant mass can then be reconstructed using these four-momenta as
\begin{equation}
       M_{X,EFO}^2 =  ( \sum_i E_{i} )^2 - ( \sum_i P_{Xi} )^2 - ( \sum_i P_{Yi} )^2 - ( \sum_i P_{Zi} )^2. 
\label{eq:mxzufo}
\end{equation}
The second method inferred $M_X$ from the measurement of the
final-state proton and the scattered positron, 
which enter into the reconstruction of the kinematic 
variables $x$ and $y$ via the double angle method. $M^2_X$ is then given by
\begin{equation}
M_{X,LPS}^2 = sy(1 - x_L - x).
\label{eq:mxlps}
\end{equation}

Since the LPS method has better resolution at higher values of $M_X$ and
the EFO measurement is better at lower values, $M_X$ was evaluated as the weighted average of the 
values obtained with the two methods. Before combining the two $M_X$ values, correction factors 
obtained from Monte Carlo studies were applied.
This resulted in a resolution on $M_X$ of about 25\% at low
$M_X$, improving to 15\% at high $M_X$.

\subsection{Boost into the $\gamma^*\P$ center-of-mass system}

The boost into the CMS of $X$
($\gamma^* \P$), Fig.~\ref{fg:feynman}(b), 
was determined from the four-momenta of the $\gamma^*$ and $\P$.
The $\gamma^*$ four-momentum was calculated from the energy and angles of the
scattered positron.
The best reconstruction of the $\P$ four-momentum was obtained by
combining the information from the LPS and the ZEUS main detector.
The $X$ and $Y$ components of the $\P$ four-momentum 
were taken to be equal to the negative of the
$X$ and $Y$ momentum components of the scattered proton measured with the LPS. 
The $Z$ component was taken to be the total $P_Z$ of all EFOs
($P_Z^{tot} = \sum_i P_{Zi} $) minus the $Z$ component of the $\gamma^*$ four-momentum.
Finally, the energy component was calculated by requiring that the $\gamma^* \P$
invariant mass be equal to the value of $M_X$ determined by the combination
of the information from the LPS and the EFOs, as described 
earlier.

The resolution on the angle between the $\gamma^* \P$
axis and the ZEUS $Z$-axis, as estimated by Monte Carlo studies, was
approximately
inversely proportional to $M_X$ and equal to 10 mrad at $M_X = 20$~GeV.
An improvement in the accuracy of the measurement by a factor of three was achieved
using the LPS compared to  a measurement without its use, i.e. substituting $p_X = p_Y = 0$.

%
%
\section{Event selection}
\label{sec:evtsel}

Deep inelastic scattering  events were selected in a manner 
similar to those of the ZEUS
$F_2$ analysis~\cite{zeusf2} by requiring:
\begin{itemize}
        \item the presence of a fully contained and
        isolated positron with energy greater than 10~GeV and 
        matched to a charged track, if within the acceptance of 
        the CTD;

        \item a value of $y$ greater than 0.05, evaluated using the Jacquet-Blondel method~\cite{jbmethod};

        \item a value of $\delta = E^{tot} - P_Z^{tot}$ in the range
        40 $< \delta <$ 65~GeV,         where
        $E^{tot}$ and $\bf{P}^{tot}$ are, respectively, the total energy and 
        momentum of the
        event in the main detector, including the scattered positron;

        \item the $Z$ coordinate of the event vertex
        within 50~cm of the nominal interaction point.
\end{itemize}
Diffractive events were then selected by requiring a well reconstructed LPS track~\cite{lps} 
carrying more than $95 \%$ of the beam energy 
($x_L > 0.95$).
The LPS track was required to pass no closer than 0.04~cm to
the HERA beampipe, and the quantity
$\delta^+ =  E^{tot} + P_Z^{tot} + 2 E_p x_L$ 
was required to be less than $2 E_p + 20$~GeV
to be consistent with longitudinal momentum conservation after
taking the LPS resolution into account. Since the results presented here were found to be
independent of $t$, no explicit cut on $t$ was applied.

To provide a sample for which the acceptance is large and uniform, to remove events 
measured with low
resolution and to reject events corresponding to exclusive production of vector mesons, 
the following kinematic cuts were applied:
\begin{itemize}
\item $70 < W < 250~{\rm GeV}$;
\item $0.0003 < \xp < 0.03$;
\item $4 < Q^2 < 150~{\rm GeV}^2$;
\item $4 < M_X < 35~{\rm GeV}$.
\end{itemize}
In addition, at least four EFOs in the system $X$ were required.
These cuts define the kinematic range for all results presented in this
paper, unless otherwise noted. The final data sample contained 
2355 events.

%
%
\section{Models of the diffractive final state}
\label{sec:MC}

The data were compared with three Monte Carlo generators based on different 
theoretical models. For all generators, hadronization was simulated using the 
Lund string model as implemented in JETSET 7.4 \cite{jetset}.
The models considered here produce 
either a $q \bar{q}$ pair or a 
$q \bar{q} g$ final state at the parton level.
However, the dynamics of the production of these states is different for each model, thus
yielding predictions that differ in their relative contributions of $q\bar{q}$ 
and $q\bar{q}g$ states, as well as in the final-state topology.
Common to all three models is that the $q\bar{q}g$-type
events dominate the final state at high masses. 
In such events, the gluon usually travels in the direction of
the Pomeron.

\subsection{Resolved Pomeron model}
\label{sec:rg}

In the resolved Pomeron approach~\cite{rg_schlein}, the exchanged particle is
assumed to have a partonic structure consisting of quarks and gluons.
A sample of resolved Pomeron events was produced with the RAPGAP 2.08/06 generator~\cite{rapgap}.
The hadronic final state is simulated in analogy to ordinary
DIS.
The Pomeron parton density functions (PDFs) used were those determined by the
H1 collaboration from their measurement of 
$d^3\sigma/d\xp dQ^2d\beta$~\cite{h1f2d4}. 

When the virtual photon in a diffractive event interacts with a quark in the Pomeron, the lowest-order final state is a $q\bar{q}$.
At ${\cal O}(\alpha_s)$,  $q \bar{q} g$ final states can be produced by the 
radiation of an extra gluon via the QCD-Compton (QCDC) process. 
In addition, 
$q \bar{q} g$ final states are produced via the boson-gluon fusion
(BGF) process between the virtual photon and a gluon in the Pomeron.
The Pomeron remnant consists of a quark in the $\qbarq$ and QCDC events 
and of a gluon in BGF events. Additional parton radiation
from the remnant partons is suppressed.
The following RAPGAP options were used:
\begin{itemize}
\item the leading-order version of the  H1 Fit 2 was used. In the fit, more than 80 \% of 
	the Pomeron momentum is carried by the gluons.
\item the final-state parton system in $\qbarq$-type 
events was simulated with limited transverse momentum with respect to the $\gamma^* \P$ axis
(intrinsic transverse momentum, $k_T$),
 randomly distributed according to $\exp{(-5.5 k_T^2)}$ ($k_T$ in GeV);
\item to avoid divergences in the matrix elements for the ${\cal O}(\alpha_s)$ processes
for massless quarks, the transverse momentum squared of any outgoing parton was required to be
larger than 3~GeV$^2$;
\item higher-order QCD radiation was simulated using initial- and final-state 
parton showers (MEPS)~\cite{meps};
\item the generator was run for light flavors and charm production (produced via the BGF process)
separately, and the final sample was obtained by mixing the two according
to their relative cross sections as predicted by RAPGAP.
\end{itemize}

The curves corresponding to this model are labeled ``RG resolved \P'' in 
later figures. The model labeled ``RG $q\bar{q}$ part only'' in later figures was obtained
by selecting events from the above sample that were not produced via the BGF process. 

\subsection{Photon dissociative model}
\label{sec:2glu}

\begin{figure}[t]
\center{ \leavevmode \epsfxsize=10cm \epsffile{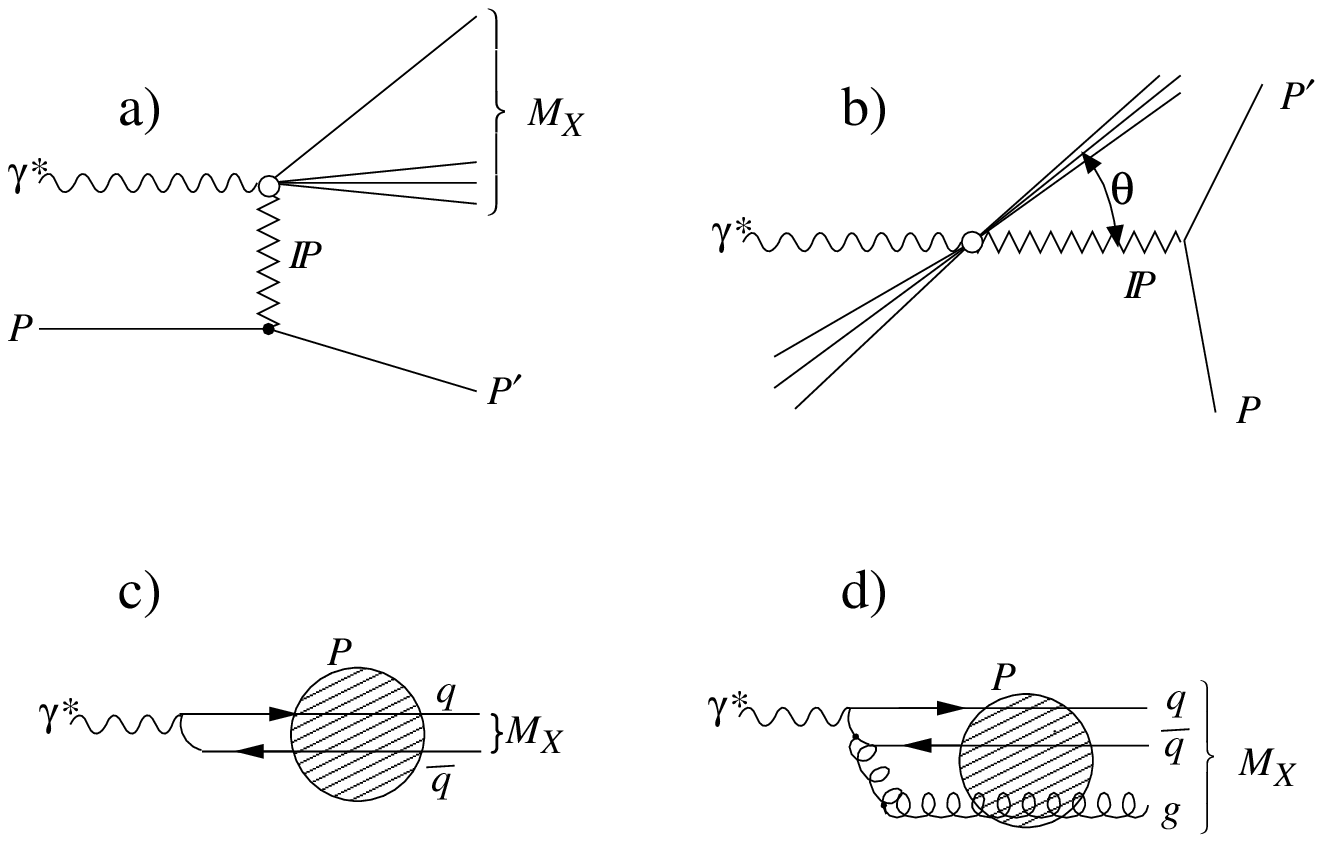} }
\caption[this space for rent]{
Different representations of diffractive DIS: (a) basic Pomeron-induced picture of 
diffraction; (b) the same process, in the $\gamma^* \P$
center-of-mass frame. The thrust angle, $\theta = \theta_{\rm Thrust}$,
 is defined as the angle between
the event axis and the $\gamma^* \P$ axis;
(c) diffraction viewed in the rest frame of the proton as the
fluctuation of the virtual photon well before the interaction with the proton
to a $q\bar q$ pair; and (d) into a $q\bar qg$ system.
}
\label{fg:feynman}
\end{figure}

In the rest frame of the proton, diffractive scattering can be viewed 
as the dissociation of the virtual photon into a $q\bar{q}$ pair 
(Fig.~\ref{fg:feynman}(c)) well before the interaction with the proton. 
For higher masses, the $q\bar{q}g$ final state (Fig.~\ref{fg:feynman}(d)) 
becomes important. The dissociated photon system couples to the proton 
by color-singlet exchange. 

The simplest realization of this color singlet is the exchange of two gluons with
opposite color charge (two-gluon exchange model).
The RIDI2.0 program~\cite{ridi} implements a two-gluon model
following the approach of Ryskin~\cite{ryskin}, where the diffractive dissociation
is treated in the framework of the leading log approximation (LLA) of perturbative QCD.
The cross section is proportional to the square of the gluon density
of the proton, which was taken from the CTEQ4M~\cite{cteq4m} parameterization
of the proton structure function with an appropriately chosen cut-off parameter 
for the transverse momentum of the final-state gluon. 
The contributions of both transversely and longitudinally polarized photons are included.
A large theoretical uncertainty in the relative contributions of $q\bar{q}$ and $q\bar{q}g$
states remains. This uncertainty is reflected in the wide range allowed for the
($\alpha_s$-dependent) $k$-factors, which determine the
relative cross sections of the $q\bar{q}$ and $q\bar{q}g$ contributions. 

Another model based on photon dissociation is that of Golec-Biernat and 
W\"usthoff~\cite{sat_theo}, in which the virtual photon
splits into a $q\bar{q}$ or $q\bar{q}g$ 
color dipole. The interaction of this
dipole with the proton can be described by an effective dipole cross section
taking the dynamics of saturation into account.
The dipole cross section was parameterized using the
HERA measurements of the DIS total cross section between 
$Q^2 = 0.1$~GeV$^2$ and $Q^2 = 100$~GeV$^2$.
The production of $q \bar{q}$  and $q\bar{q}g$ final states is calculated using these parameters.
The relative contribution of $q \bar{q}$ and
$q \bar{q} g$ events is fixed by QCD color factors.
This model is implemented in the SATRAP Monte Carlo generator~\cite{satrap}
which uses the additional approximation of strong ordering in the transverse momenta: 
$k_T(g) \ll k_T(q, \bar{q})$. 
In the course of the ZEUS analysis of diffractive 3-jet events~\cite{3jet}, it was found 
that the modeling
of higher-order QCD processes was inadequately implemented in SATRAP. Most notably, no initial-state
parton cascades were included, and the final-state QCD radiation from the gluon in the 
dominant $q\bar{q}g$ contribution was suppressed. A new implementation of higher-order
QCD processes in SATRAP was carried out~\cite{sr-cdm1, sr-cdm2}, in which the
color-dipole model (CDM)
was implemented in a similar fashion to that in RAPGAP. This model is referred to
as SATRAP-CDM.

\subsection{JETSET}

If the diffractive DIS hadronic system, $X$,
 is produced by a virtual photon splitting into a $q\bar{q}$ pair, its properties would be expected
to be similar to those of the hadronic system produced by
$e^+ e^-$ annihilation at a CMS energy $\sqrt{s} = M_X$. 
The $e^+e^-$ final state, as simulated by the JETSET program~\cite{jetset},
 was used as a baseline 
to which both the data and the
diffractive Monte Carlo generators were compared.
JETSET is known to
describe accurately many details of the final state
in $e^+ e^-$ collisions. 
 
%
%
\section{Data correction and background}
\label{sec:signal_mc}

\begin{figure}[b] 
\center{ \leavevmode \epsfxsize=15cm \epsffile{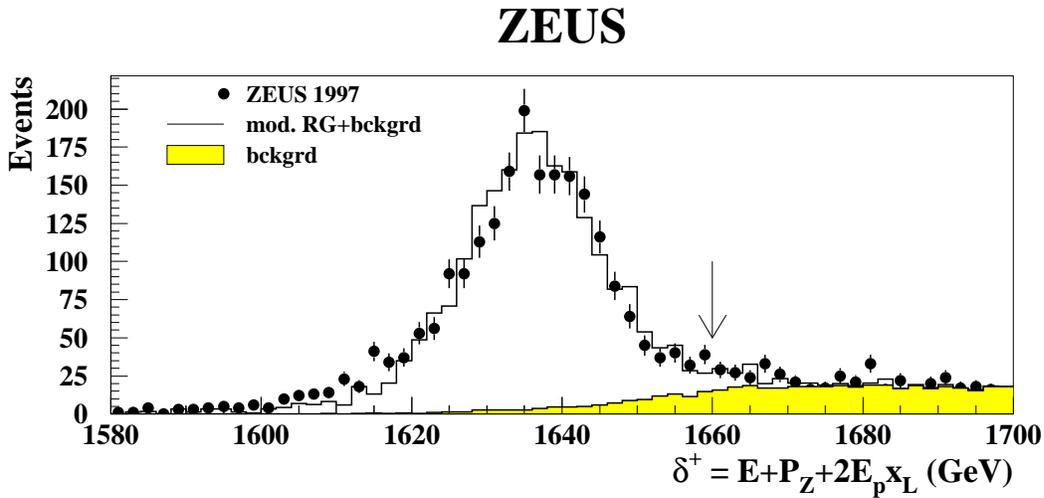} }
\caption[this space for rent]{
Normalization of the background. The quantity 
$\delta^+ = \Sigma (E_i + P_{Zi}) + 2 E_p x_L$ is shown
for data (points, with statistical error only), 
background (shaded histogram) and modified RAPGAP plus
background (solid histogram). The normalization was obtained as described in the text.
The vertical arrow indicates the cut below which events were accepted for the analysis.
}
\label{fg:epluspz}
\end{figure}

\begin{figure}[p]
\center{ \leavevmode \epsfxsize=15cm \epsffile{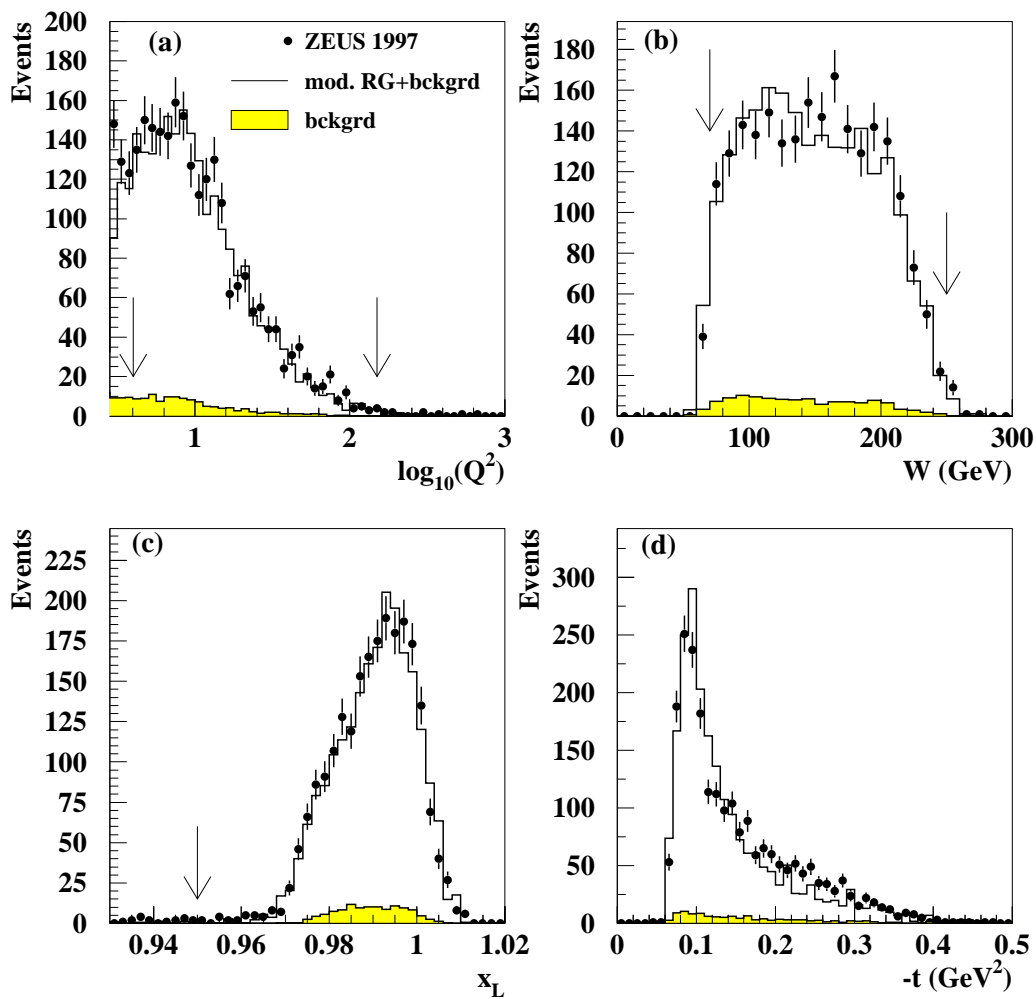} }
\caption[this space for rent]{
A comparison of data and Monte Carlo simulation for
distributions of the kinematic variables of the diffractive DIS
sample: (a) $Q^2$, (b) $W$, (c) $x_L$, and (d) $-t$. 
The data are displayed as points, with statistical errors only, and are 
compared to the modified RAPGAP plus background
(solid histogram) and background alone (shaded histogram).
The normalization was obtained as described in the text.
Vertical arrows in (a) and (b) indicate the values of $Q^2$ and $W$
between which events were selected for
this analysis. Events with $x_L$ values below that indicated by the arrow in (c)
were rejected. No cut on $t$ was imposed.
}
\label{fg:mcvsdata}
\end{figure}

\begin{figure}[p]
\center{ \leavevmode \epsfxsize=15cm \epsffile{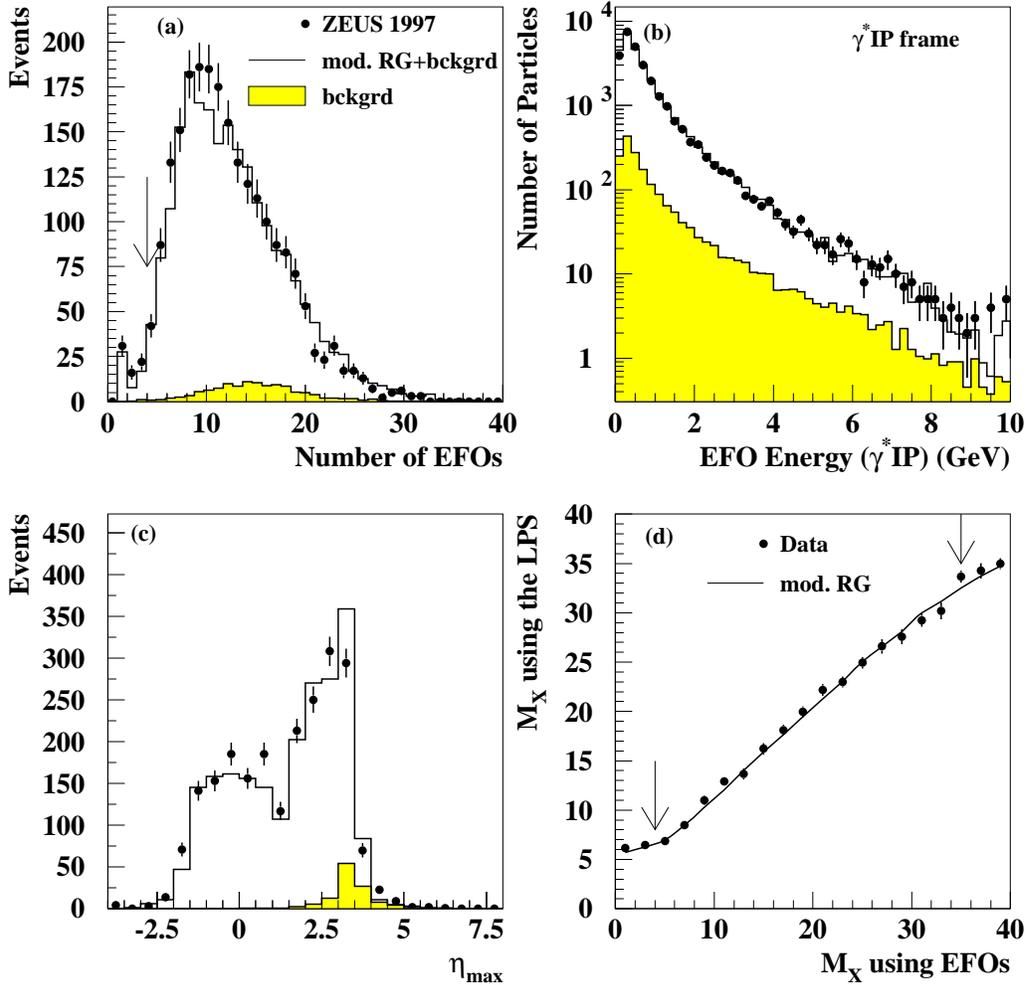} }
\caption[this space for rent]{
A comparison of data and Monte Carlo simulation for
distributions related to the properties of the reconstructed hadronic
final state: (a) the number of EFOs; 
(b) the energy spectrum of the EFOs 
in the $\gamma^* \P$ frame; (c) $\eta_{max}$; and
(d) the average invariant-mass $M_X$ measured with the 
LPS alone versus $M_X$ determined from the EFOs. 
The data are displayed as points, with statistical errors only, and are 
compared to the modified RAPGAP plus background
(solid histogram), background alone (shaded histogram) or modified RAPGAP alone (curve in (d)).
The vertical arrow in (a) indicates the cut above which events were selected for this analysis.
The region of $M_X$ considered is shown by the arrows in (d). No cuts were imposed on the
particle energy and $\eta_{max}$.
}
\label{fg:mxdist}
\end{figure}

\begin{figure}[p]
\center{ \leavevmode \epsfxsize=15cm \epsffile{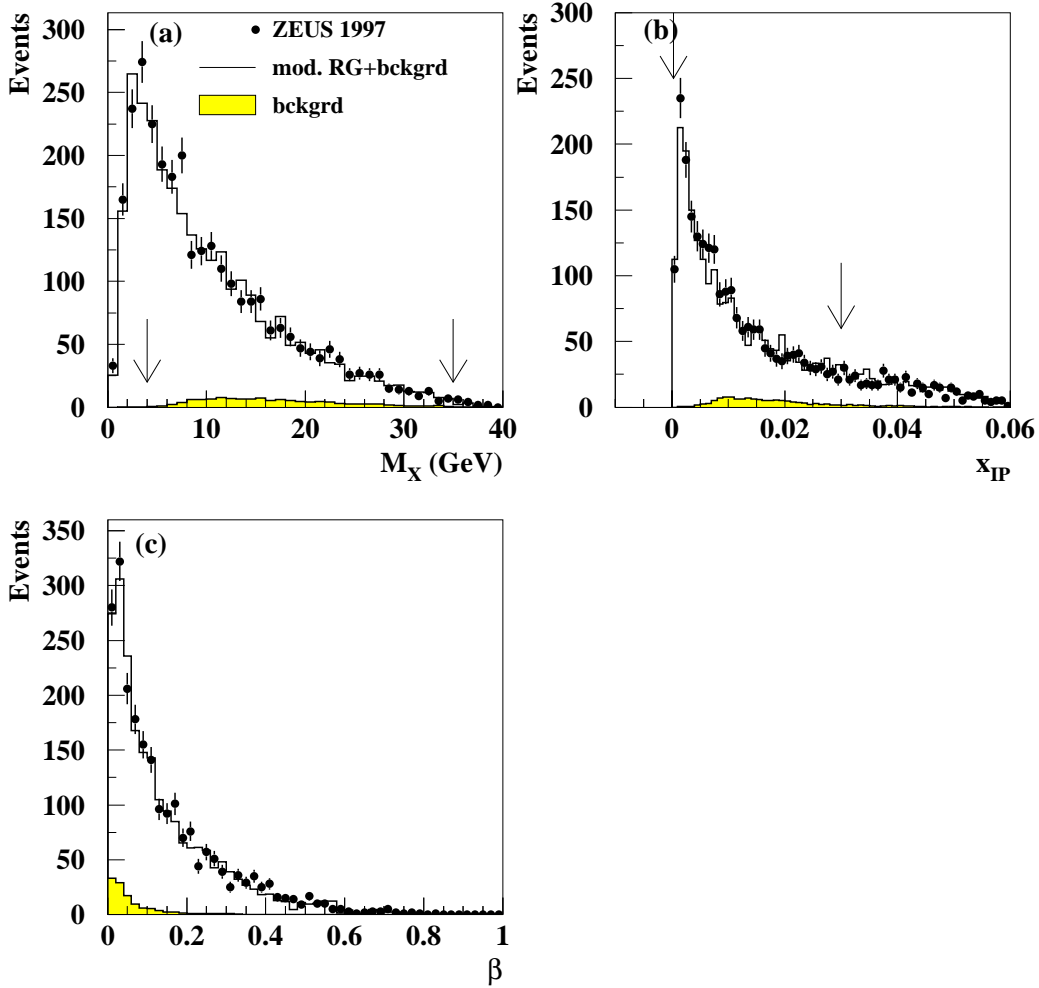} }
\caption[this space for rent]{
A comparison of data and Monte Carlo simulation for
measured distributions of (a) $M_X$, (b) $\xp$ and (c) $\beta$.
The data are displayed as points, with statistical errors only, and are 
compared to the modified RAPGAP plus background
(solid histogram). The background estimate is shown by the 
shaded histogram.
The vertical arrows in (a) and (b) indicate 
the values of $M_X$ and $\xp$ between which events were selected for this analysis. 
No cut on $\beta$
was imposed.
}
\label{fg:mxxpombeta}
\end{figure}

Monte Carlo simulations were used to correct the data for the resolution and acceptance of
the main detector and the LPS,
and to estimate the size and influence of the background.
To estimate the model dependence of these corrections, two different generators
were used: RAPGAP~\cite{rapgap} and RIDI~\cite{ridi}.
RAPGAP events were generated with  the 
H1 QCD Fit~2~\cite{h1f2d4} for the \P~structure function, as
described in Section~\ref{sec:rg}, except that CDM
rather then MEPS was used for higher-order initial- and final-state QCD radiation. 
This model gives a poor description of the hadronic
final state as well as of some kinematic variables as measured at the
 detector level. To obtain a 
sample which gives a good description of all observed distributions, the RAPGAP generator 
was reweighted to reproduce the data. The corresponding distributions are labeled ``mod. RG''
in Figs. \ref{fg:epluspz} to \ref{fg:mxxpombeta}.
A similar reweighting procedure was carried out for RIDI.
Both samples were passed through a GEANT 3.13~\cite{geant}
simulation of the ZEUS detector, subjected to the same
trigger requirements as the data and processed by the
same reconstruction programs. 

In addition, changes in the HERA beamline parameters during the running period and 
the finite resolution of the motors that determine the position of the LPS, neither of
which was simulated in the Monte Carlo,
were taken into account by reweighting and smearing the
simulated $x_L$ distribution.


The primary source of background in this analysis is 
the accidental overlap of a DIS event with an unrelated beam-halo proton
measured in the LPS. This contribution was assumed to be constant over
the running period.
To estimate its size, a sample of such background events was studied.
Background
protons were identified
by selecting DIS events with $E^{tot} + P_Z^{tot} > 100$~GeV (using only CAL)
and an LPS track with $x_L > 0.9$. 
From energy conservation allowing for detector
resolution, such events must 
result from an accidental overlap.
The LPS information from these events was then combined with a sample of
non-diffractive DIS Monte Carlo events generated using DJANGOH~\cite{djangoh}.
The resulting sample was normalized
to match the upper tail
of the $\delta^+$
distribution for the data shown in Fig.~\ref{fg:epluspz}.
With this method, the level of background contamination, after all selection cuts, was 
estimated to be 5.2\%. Its contribution was statistically subtracted in all
results presented below.

Figure~\ref{fg:mcvsdata} shows the distributions of $Q^2$,  
$W$, $x_L$ and $-t$ for events 
which pass all selection cuts except the ones imposed on the variables shown 
(indicated by the arrows).  Also shown in the figure are the 
distributions from
the modified RAPGAP simulation plus background
which was normalized 
so that the weighted sum of the modified RAPGAP 
and the background Monte Carlo events describes the $Q^2$ distribution
in the selected range, see Fig.~\ref{fg:mcvsdata} (a).
The kinematic distributions of the data, with the exception of  the $t$ distribution, are well
described by the modified RAPGAP  Monte Carlo sample in combination with the 
background sample.


Figure~\ref{fg:mxdist} (a-c) shows some of the properties of the hadronic final 
state: the distribution of the number of EFOs, the EFO energy in the CMS
frame of the
$\gamma^*\P$ system and $\eta_{max}$, where $\eta_{max}$ is the 
pseudorapidity of
the most-forward calorimeter cluster of energy greater than 400~MeV.
The data are well
described by the modified RAPGAP  sample in combination with the background sample. Figure~\ref{fg:mxdist} (d) shows the average $M_X$ reconstructed using the LPS (Eq.~(\ref{eq:mxlps})) as a function
of the invariant
mass calculated using EFOs (Eq.~(\ref{eq:mxzufo})).
The correlation seen in the data is well reproduced
by the Monte Carlo simulation. The fact that the correlation is approximately linear 
in the selected
mass range confirms that the final state is well contained in the ZEUS
detector.

Figure~\ref{fg:mxxpombeta} shows the distributions of $M_X$, $\xp$ and $\beta$. After
the reweighting procedure, all data distributions are well
described by the modified RAPGAP Monte Carlo sample.
This reweighting, which mainly affected the low $M_X$ and
high $\xp$ regions, had little effect on the $\beta$ distribution
which, although peaking at low values, is well
described by the relatively flat Pomeron PDFs used in RAPGAP.
RIDI (not shown in Figs.~\ref{fg:epluspz} to \ref{fg:mxxpombeta}) also reproduces
the data satisfactorily.



%
%

\section{Systematic uncertainties}
\label{sec:SYS}

The systematic uncertainties were obtained by studying the sensitivity of
the results to the uncertainties in the understanding of the ZEUS
detector response
and by modifying the analysis procedures as listed below.

Uncertainties related to the understanding of the ZEUS main detector include:
\begin{itemize}
\item the uncertainty on the absolute energy scales in each 
	major section of the calorimeter
	(FCAL, BCAL, and RCAL), which for this analysis was 
	understood to a level of 3\% for hadrons and 2\% for the scattered positron;
\item the uncertainty on the survey measurements of the positions
	of the  major sections of the calorimeter with respect to each other and to the HERA
	beamline, which are accurate to 1--2~mm.
\end{itemize}

Uncertainties related to the LPS include:
\begin{itemize}
\item  possible shifts in $x_L$, $p_X$, and $p_Y$ of $\pm 0.003$, 
$\pm 10$~MeV, and $\pm 50$~MeV, respectively, as determined from an analysis
of elastic $\rho$ photoproduction ($ep \rightarrow e\rho p$); 
\item the uncertainty on the beampipe position,
as determined by alignment studies, was taken into account 
by changing the cut on the distance of closest approach of the proton track to the beampipe
by $\pm 400$~$\mu {\rm m}$;
\item the uncertainty in the Monte Carlo simulation of the proton track reconstruction
was taken into account by applying tighter cuts on the quality of the Monte Carlo track.
\end{itemize}

To check for the effect of possible particle losses into the forward beampipe, and to
account for the differences in modeling this forward region in the Monte Carlo generators,
the energy deposited in the inner ring of the FCAL, which 
covers approximately the pseudorapidity range of $2.7 < \eta < 3.9$,
was scaled by $\pm 25 \%$, as suggested by MC studies.

The normalization of the background from an accidental overlap of a DIS event with an 
unrelated proton was estimated using the $\delta^+$ distribution of Fig.~\ref{fg:epluspz}.
The uncertainty caused by the background subtraction was determined by
changing the background normalization by $\pm 50 \%$.

To estimate the model dependence of the results, the data were also corrected using 
RIDI. The difference between the RIDI and modified RAPGAP results was taken as an 
estimate of the model uncertainty.  
This uncertainty was assumed to be symmetric with respect to the nominal results obtained
using the modified RAPGAP sample. 

The largest contributions to the systematic uncertainty typically originate from the 
model dependence and the uncertainty in the $x_L$ reconstruction.
All systematic
uncertainties were assumed to be independent and were 
calculated
separately for positive and negative variations with respect to the nominal value. 
The total positive and negative systematic uncertainties were calculated as the 
corresponding sums in quadrature.

%
%
\section{Global event-shape variables}
\label{sec:gevt}

The results presented here are corrected to the hadron level in the 
kinematic range defined in Section~\ref{sec:evtsel} with at least four stable
particles in the final state.
A particle is considered
stable if its lifetime is longer than $3 \times 10^{-10}$~s. 
If the lifetime is 
shorter than $3 \times 10^{-10}$~s, the daughter particles 
(with lifetime $> 3 \times 10^{-10}$~s) are included. This corresponds to the 
final-state definition used by $e^+e^-$ experiments~\cite{eedata}.

\subsection{Thrust and sphericity}

The event-shape-variable thrust ($T$) is calculated by determining 
the unit vector $\bf{\hat{n}}$ which maximizes
\begin{eqnarray}
T({\bf \hat{n}}) = { \frac{\sum_i | {\bf \hat{n}} \cdot {\bf p}_i |} 
                    {\sum_i | {\bf p}_i |}},
\label{eq:thrust}
\nonumber
\end{eqnarray}
where ${\bf p}_i$ is the three-momentum of a final-state particle and
the sum is over all particles belonging to the system under study.
The resulting axis ${\bf \hat{n}}$ is called the thrust axis
and $T({\bf \hat{n}})$ is the thrust value. For collimated two-jet
events, the value of $T$ approaches 1, while events with an isotropic
shape yield values close to 0.5.

The sphericity ($S$) is defined as
\begin{eqnarray}
S = \frac{3}{2} (\lambda_2 + \lambda_3), \label{eq:s}
\nonumber
\end{eqnarray}
where $\lambda_2$ and $\lambda_3$ are the two smallest eigenvalues 
(corresponding to orthogonal eigenvectors) of the
sphericity tensor
\begin{eqnarray}
S^{\alpha \beta} = { \frac{\sum_i p_i^\alpha p_i^\beta} 
                          {\sum_i | {\bf p}_i |^2} }
        \quad \alpha,\beta = x,y,z.
\label{eq:s_tensor}
\end{eqnarray}
For isotropic events, $S$ approaches 1, and for collimated two-jet events,
$S$ is close to 0.

\begin{figure}[p] 
\center{ \leavevmode \epsfxsize=17cm \epsffile{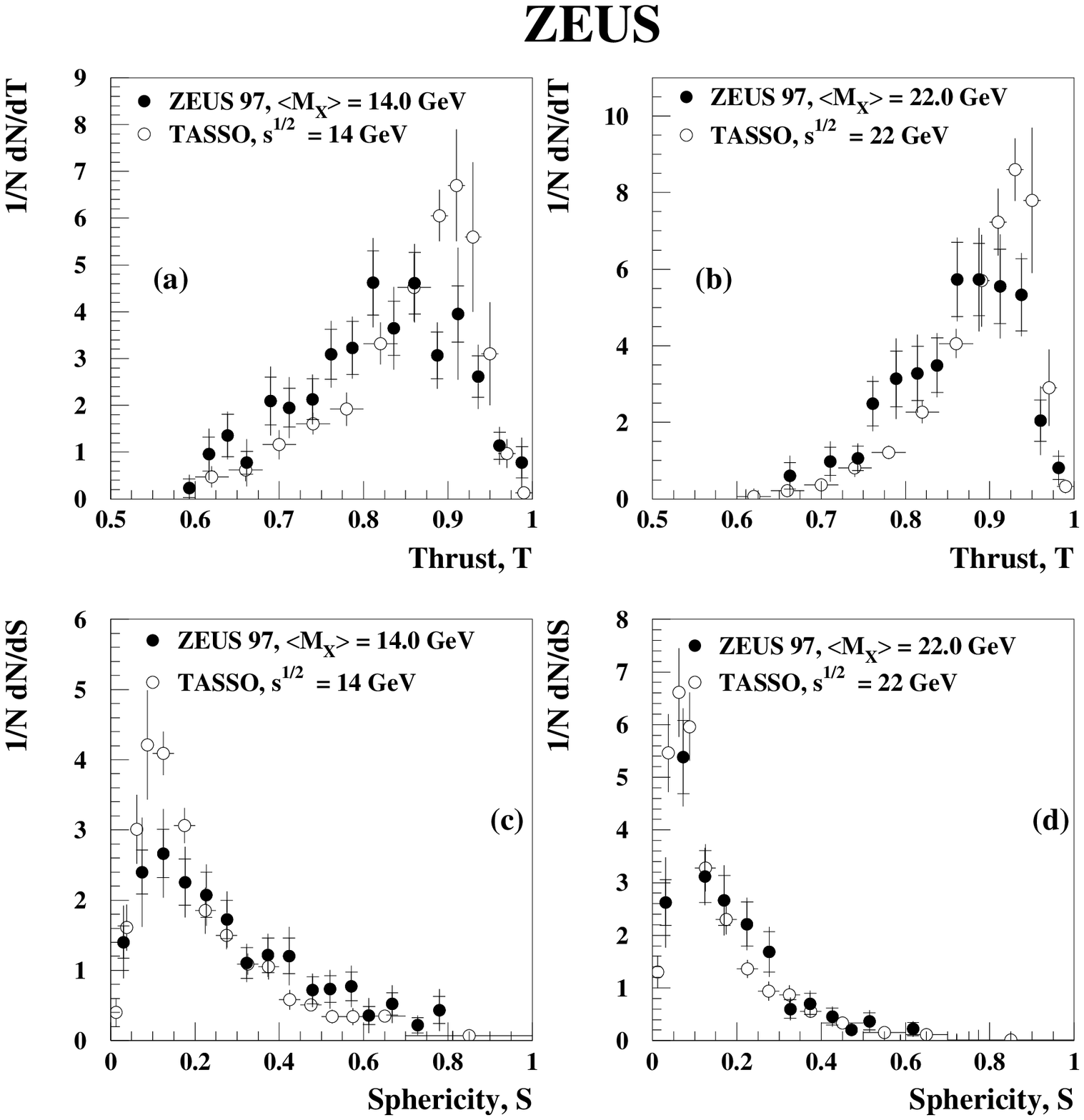} }
\caption[this space for rent]{
  Thrust, $T$, and sphericity, $S$, distributions of the diffractive DIS
  hadronic final state (filled circles) 
  compared to distributions measured in
  $e^{+}e^{-}$ collisions by the TASSO collaboration~\cite{eedata} 
  (open circles) at
  the same CMS energies of the system $X$. 
  In (a) and (c), results for
  $11 < M_X < 17.8$ GeV (\avM $ = 14.0$ GeV) are compared to measurements in
  $e^{+}e^{-}$ collisions at $\sqrt{s} = 14$ GeV; in (b) and (d),
  the results for $17.8 < M_X < 27.7$ GeV (\avM $= 22.0$ GeV) are compared to
  $e^{+}e^{-}$ results at $\sqrt{s} = 22$ GeV. The inner error bars
  show statistical uncertainties only; the outer bars show the statistical and
  systematic uncertainties added in quadrature.}
\label{fg:tsdist}
\end{figure}

The sphericity and thrust distributions in the CMS of the final
state in diffractive DIS are shown in Fig.~\ref{fg:tsdist} for two bins
of $M_X$, and are compared
to the ones observed in $e^+ e^-$ annihilation from the TASSO collaboration~\cite{eedata}
at $\sqrt{s} = $\avM.  The diffractive events show thrust (sphericity)
distributions which are broader and shifted to lower (higher) values compared to
the $e^+e^-$ results, indicating that they are more isotropic.
The diffractive events become more collimated as
$M_X$ increases, a trend also observed in $e^+e^-$ annihilation.   
If the virtual photon in diffractive DIS fluctuates only into a $q\bar{q}$ state, 
the resulting hadronic final state should
develop in a manner similar to $e^+ e^- \to q\bar{q}$
at a CMS energy $\sqrt{s} = $\avM.
Deviations from this $e^+ e^-$-type of behavior are expected, however,  since there
is a significant contribution from the
$\qbarqg$ diagram in the color field of the proton (Fig.~\ref{fg:feynman}(d)), 
which is not present in 
$e^+ e^-$ collisions, where gluons can only be produced via higher-order QCD
radiation.

\begin{figure}[p]
\center{ \leavevmode \epsfxsize=17cm \epsffile{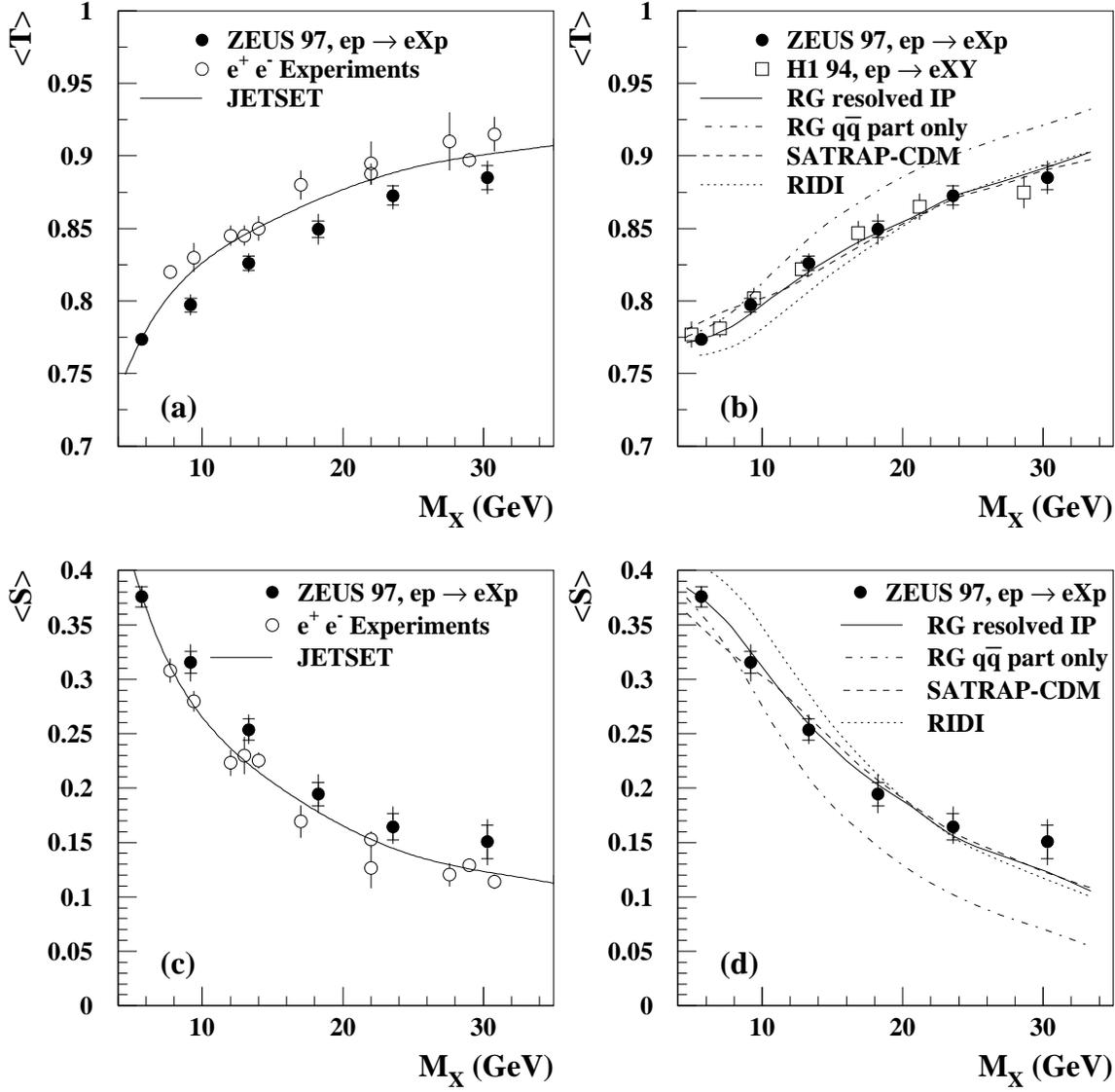} }
\caption[this space for rent]{
  Average thrust \avT~and sphericity \avS~of the diffractive DIS
  hadronic final state as a function of $M_X$. Shown for comparison
  are results from $e^+ e^-$ collisions and predictions from three
  Monte Carlo generators, RAPGAP (solid), RAPGAP $q\bar{q}$ only (dot-dashed),
SATRAP-CDM (dashed) and RIDI (dotted). 
  Also shown in (b) are thrust values from the H1
  Collaboration for diffractive DIS events tagged using rapidity gaps
  \cite{h1eventshape}.  The inner error bars show statistical uncertainties
  only; the outer bars are the statistical and systematic uncertainties
  added in quadrature.
 }
\label{fg:ts}
\end{figure}

Figures~\ref{fg:ts} (a) and (c) show 
the average values of thrust, \avT, and sphericity, \avS,
measured in six bins of $M_X$.
The value of \avT~increases and \avS~decreases with increasing
$M_X$ with slopes similar to those found in the $e^+e^-$ data \cite{eedata}. 
On average, the diffractive events are less
collimated than $e^+e^-$ events at a similar CMS energy.

Figures~\ref{fg:ts} (b) and (d) show the same data, but now compared to
several Monte Carlo models\footnote{Henceforth, all comparisons of the data are made to
MC models that have not been reweighted in the manner described in Sec.~\ref{sec:signal_mc}.}.
Also plotted in Fig.~\ref{fg:ts} (b) are the average thrust values measured by the H1 collaboration
using a diffractive DIS sample tagged with rapidity
gaps~\cite{h1eventshape}. 
The present measurements
are in good agreement with the H1 results which have been corrected to the 
full phase space by a MC simulation.
The results are also consistent with those presented in a previous ZEUS 
publication~\cite{zeuseventshape}, which 
were obtained for a small range in $\eta$.

The Monte Carlo models of diffractive scattering reproduce the
general trend of the data.
RAPGAP and SATRAP-CDM give a good description of \avT~over almost the full $M_X$ range, 
while RIDI fails for $M_X~\leq~20$~GeV.
The diffractive models produce events more isotropic than the ones 
generated with JETSET and measured in
$e^+e^-$ annihilation. 
This is mainly due to the inclusion of $q\bar{q}g$ final states.
Figures~\ref{fg:ts} (b) and (d) also show  the RAPGAP contribution for the
$q\bar{q}$ diagram alone. These curves demonstrate the need for the $q\bar{q}g$
contribution, especially at the higher $M_X$ values.

\begin{figure}[t] 
\center{ \leavevmode \epsfxsize=17cm \epsffile{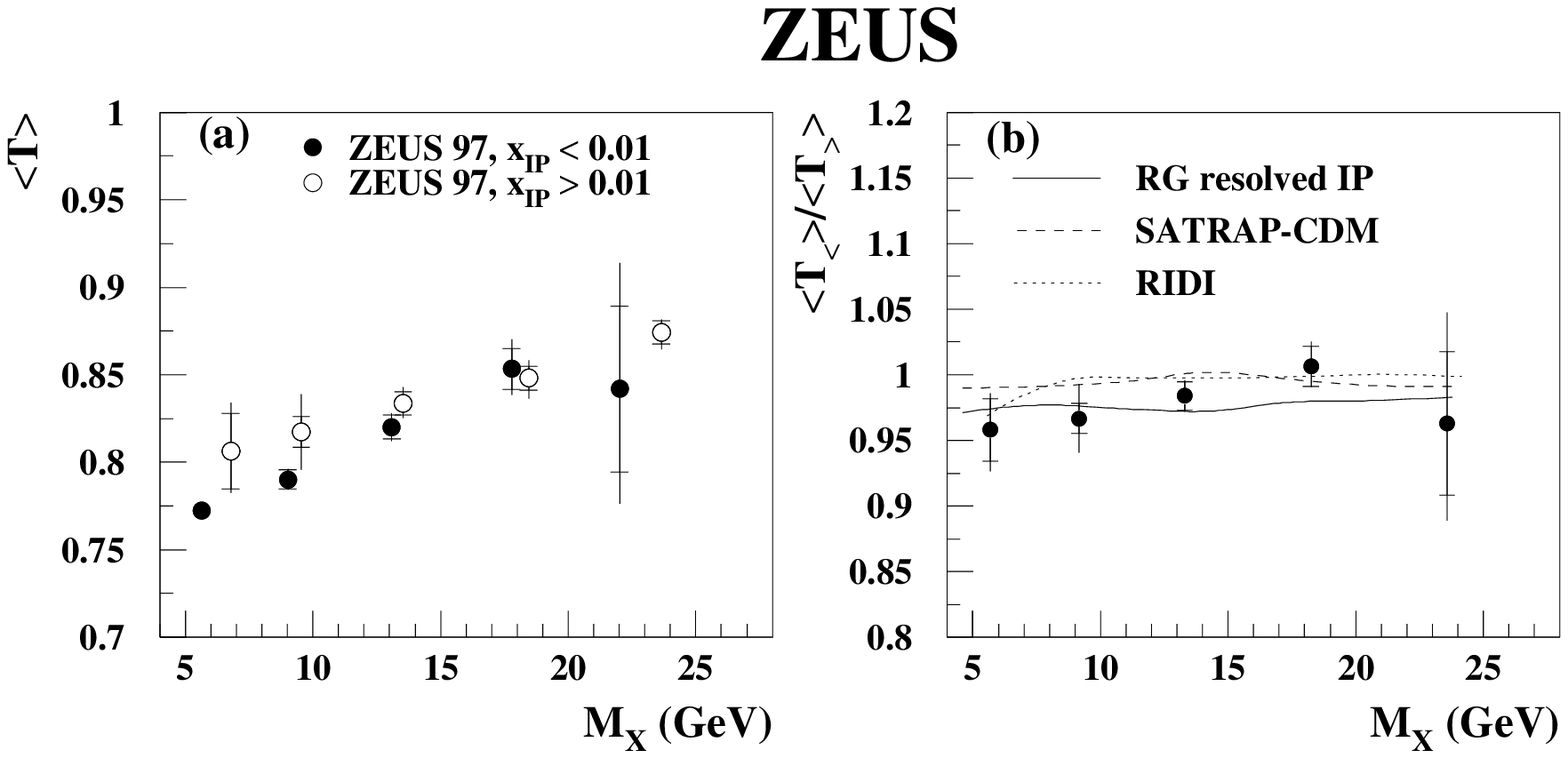} }
\caption[this space for rent]{
  (a) Average thrust, \avT, of the diffractive DIS
  hadronic final state as a function of $M_X$ for two
  different samples of events: 
  $\xp < 0.01$ (solid circles) and $\xp > 0.01$ (open circles). 
  (b) Ratio of average thrust values $T_< = T(\xp < 0.01)$ and  
     $T_> = T(\xp > 0.01)$ as a function of $M_X$ for data (solid circles), and 
     for predictions from RAPGAP (solid curve), SATRAP-CDM (dashed curve), and RIDI (dotted curve).  
     The inner error bars show statistical uncertainties only; the outer
  bars are the statistical and systematic uncertainties added in
  quadrature. \\
}
\label{fg:ts_xpom}
\end{figure}

The final state measured in terms of thrust and sphericity 
has no strong dependence on any kinematic variable other than $M_X$.
Figure~\ref{fg:ts_xpom} shows \avT~as a function of $M_X$ for $\xp < 0.01$,
where Pomeron exchange dominates,
and for $\xp > 0.01$, where Reggeon exchange may become more important.
No significant differences are observed and all models give a reasonable
description of the data.
The data were also split into two samples of $Q^2$, $t$ and $x$; 
the dependence on these variables was less pronounced than that for $\xp$.


\subsection{Transverse momenta in and out of the event plane}
\label{sec:ptinout}

The shape of the system $X$ was also studied in terms of 
two more variables, $P^2_{T,in}$ and $P^2_{T,out}$, which measure the
transverse momentum {\it in} and {\it out} of the event plane. The event plane is  defined 
by the eigenvectors of the sphericity tensor associated with the two largest
eigenvalues, $\lambda_1$ and $\lambda_2$ (see Eq.~(\ref{eq:s_tensor})).
These transverse momenta have also been studied in $e^+ e^-$ experiments, since they are
sensitive to gluon bremsstrahlung.
They are evaluated by multiplying the average momentum squared
of the $n$ particles in the event by the two smallest sphericity eigenvalues 
$\lambda_2$ and $\lambda_3$, respectively:
\begin{eqnarray}
P^2_{T,in} & = & \frac{\sum_i p_i^2}{n} \lambda_2, \\
P^2_{T,out} & = & \frac{\sum_i p_i^2}{n} \lambda_3 ,
\end{eqnarray} 
where $p_i^2$ is the squared momentum of the $i$th particle
 in the event. By definition, $P^2_{T,in} > P^2_{T,out}$.
The observation of a
difference in the $M_X$ behavior of these two variables could be 
explained by the presence of events of planar shape, such as events with 
three partons in the final state.

\begin{figure}[b]
\center{ \leavevmode \epsfxsize=15cm \epsffile{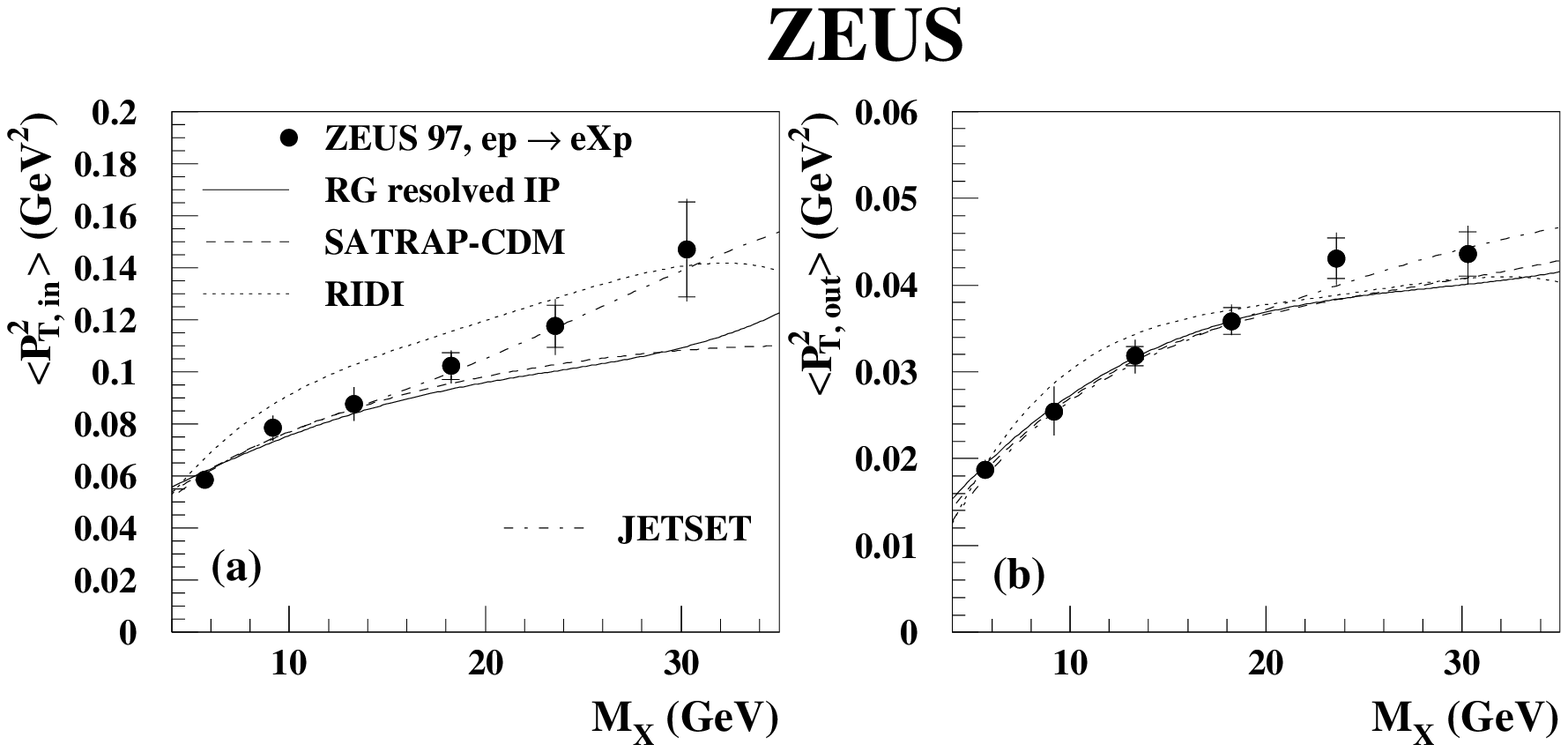} }
\caption[this space for rent]{
  (a) Average $P^2_{T,in}$ and (b) $P^2_{T,out}$ 
 of the diffractive DIS
  hadronic final state as a function of $M_X$. Shown for comparison
  are predictions from three diffractive
  Monte Carlo generators, RAPGAP (solid curve), SATRAP-CDM (dashed curve), and RIDI (dotted curve) 
  together with $e^+e^-$ results as simulated by JETSET (dashed-dotted  curve).
  The inner error bars show statistical uncertainties
  only; the outer bars are the statistical and systematic uncertainties
  added in quadrature.
  \vspace{2ex}}
\label{fg:ptinout}
\end{figure}

Figures~\ref{fg:ptinout} (a) and (b) show the average values of $P^2_{T,in}$ and $P^2_{T,out}$
for the diffractive DIS sample as a function of $M_X$. 
On average, $P^2_{T,in}$ is about a factor of three larger than  
$P^2_{T,out}$. The dependence of $P^2_{T,in}$ on $M_X$ is almost linear over the
range studied, while the $M_X$ dependence of $P^2_{T,out}$ becomes weaker as $M_X$ increases,  
indicating that the diffractive events become more planar.
Figure~\ref{fg:ptinout} also shows the predictions of the diffractive DIS 
generators and JETSET.
The RAPGAP and SATRAP-CDM samples give a good
description of $P^2_{T,in}$ for low masses, but underestimate its value for
$M_X~\sgeq~15$~GeV. RIDI overestimates $P^2_{T,in}$ for most of the measured  $M_X$ region. 
Transverse momentum out of the event plane can be produced by the fragmentation process. 
While RAPGAP and SATRAP-CDM give a reasonable description of the data, RIDI
slightly overestimates the data at low $M_X$ values.
The transverse momentum distributions in and out of the event plane 
predicted by JETSET are in excellent agreement with the 
diffractive scattering data.

In summary,
the diffractive hadronic final state shows trends in 
\avT~and \avS~different from those observed in $e^+e^-$ annihilation.
These differences can be 
explained by the presence of $q\bar{q}g$ events from BGF (RAPGAP) or from
photon dissociation (SATRAP-CDM, RIDI) which are absent in $e^+e^-$ annihilation
 and JETSET,
where the only source of gluons in the final state is from QCD radiation.
The transverse momentum out of the event plane obtained from diffractive Monte Carlo models 
and JETSET reproduces the data, indicating that
the hadronization process is consistently modeled and is similar in 
diffractive DIS and $e^+e^-$ annihilation.

%
%
\section{Thrust axis orientation}
\label{sec:tht_t}

The orientation of the entire hadronic final state, $X$,
with respect to the $\gamma^* \P$ axis can be
studied by measuring the angle $\theta_{\rm Thrust}$ (as shown in Fig.~\ref{fg:feynman} (b)) 
between this
axis and the thrust axis.

The values of $\theta_{\rm Thrust}$ 
and the transverse momentum, $P_{\rm T}^{\rm Thrust}$, relative 
to the thrust axis, defined by
\begin{eqnarray}
P_{\rm T}^{\rm Thrust} = \frac{M_X}{2} \sin{\theta_{\rm Thrust}},
\label{eq:ptt}
\end{eqnarray}
are sensitive to various sources of transverse momentum. 
These include the intrinsic transverse momentum, $k_T$, of the partons in the proton, 
which was found to be 1.69$\pm 0.18 ^{+0.18}_{-0.20}$~GeV in a recent ZEUS publication~\cite{promptPhoton},
intrinsic transverse momentum of the partons in the Pomeron, and contributions
of hard processes, e.g. final-state gluon emission~\cite{hera-workshop:kt}.

\begin{figure}[t]
\center{ \leavevmode \epsfxsize=15cm \epsffile{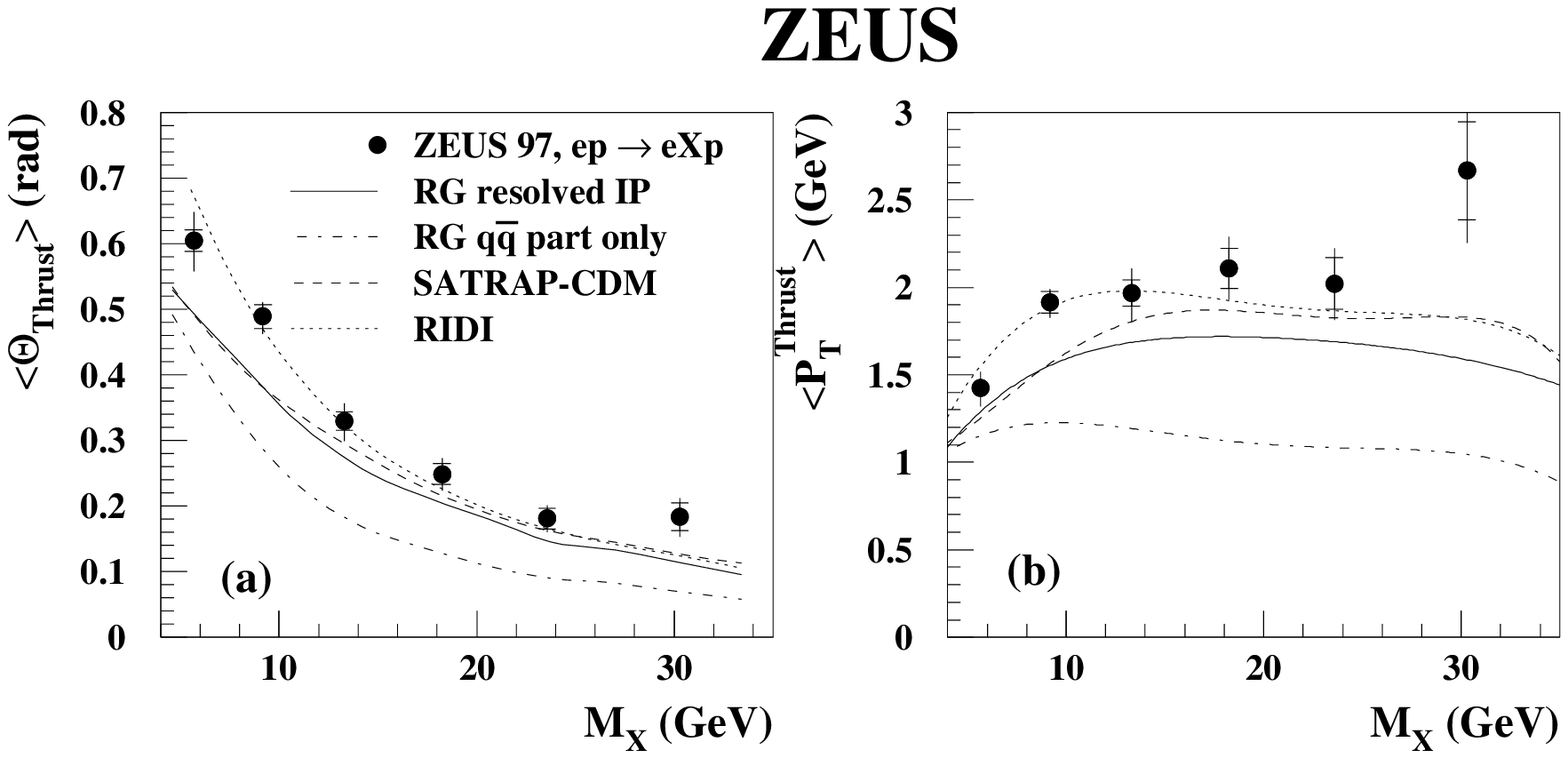} }
\caption[this space for rent]{
(a) The average value of the angle $\theta_{\rm Thrust}$ of the thrust 
axis in the $\gamma^* \P$
frame  and (b) the corresponding transverse momentum $p_{\rm T}^{\rm Thrust}$ 
compared to the prediction of RAPGAP (solid curve), $q\bar{q}$ part of 
RAPGAP (dot-dashed), SATRAP-CDM (dashed curve) and 
RIDI (dotted curve).
The inner error bars show statistical uncertainties
only; the outer bars show the statistical and systematic uncertainties
added in quadrature.
}
\label{fg:theta}
\end{figure}

Figure~\ref{fg:theta} (a) shows the average thrust angle \avthtt~as a function
of $M_X$. The angle is largest at small masses and falls off steeply as $M_X$
increases, indicating that the event becomes more aligned with the $\gamma^* \P$ axis
as the CMS energy increases.
This general trend is described by all Monte Carlo models shown, but only
RIDI produces a thrust angle as large as that observed in the data.
SATRAP-CDM and RAPGAP fall about equally short of the observed angles.
Given that the hadronization is well described by all Monte Carlo models 
(see Section~\ref{sec:ptinout}), 
the measurement of \avthtt~implies that additional sources of transverse momentum 
as discussed above must be important.
Indeed, the approach implemented in RIDI
favors the production of partons with relatively large transverse momenta,
typically of the order of 1 GeV~\cite{ridi}.

Figure~\ref{fg:theta} (b) shows that the average transverse momentum, \avptt, 
produced in diffractive scattering is almost independent
of $M_X$ for \mbox{$M_X > 10$~GeV}, with an average value of about 2~GeV.
In this region of $M_X$, the independence is correctly reproduced by all 
Monte Carlo models, but the RAPGAP prediction is too low.
The need for the $q\bar{q}g$ contribution is illustrated by the curve showing only
that part of the RAPGAP prediction corresponding to the $q\bar{q}$ final state.

%
%

\section{Energy flow}
\label{sec:eflow}

\begin{figure}[p]
\center{ \leavevmode \epsfxsize=17cm \epsffile{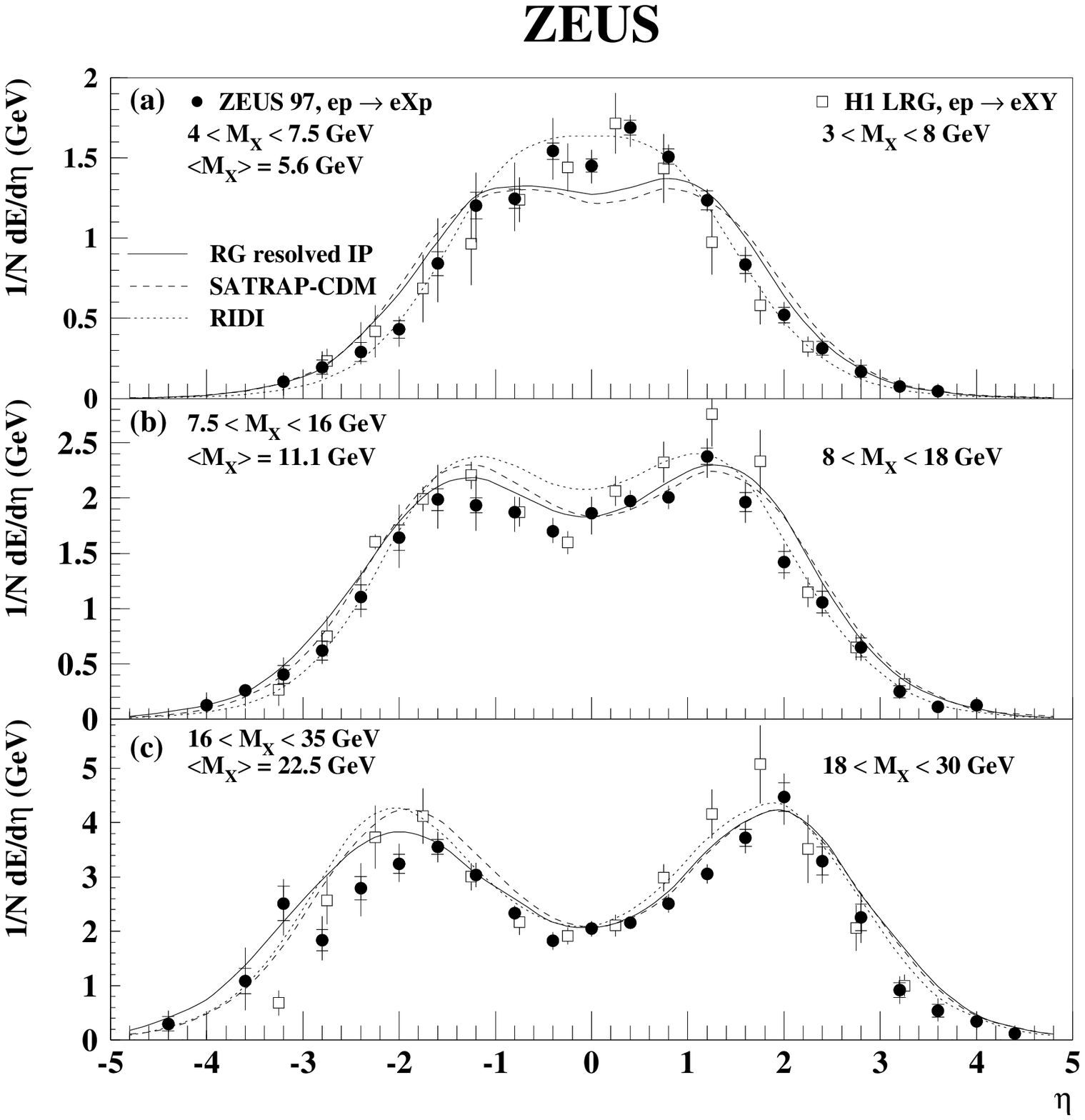} }
\center{ \leavevmode \epsfxsize=10cm \epsffile{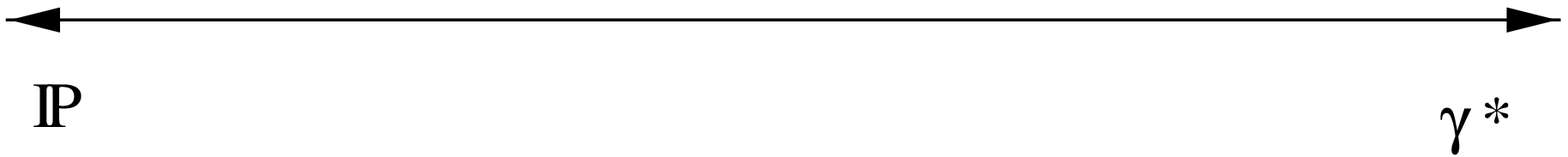} }
\caption[this space for rent]{
  The energy flow as a function of pseudorapidity, $\eta$, in the $\gamma^* \P$-CMS, 
  for various ranges of $M_X$.
  The data are shown as solid points, while the Monte
  Carlo predictions from RAPGAP, SATRAP-CDM and RIDI are shown as solid, 
  dashed and dotted curves, respectively.
  Positive $\eta$ is in the direction of the
  $\gamma^*$. The inner error bars show statistical uncertainties only; the
  outer bars show the statistical and systematic uncertainties added in
  quadrature.
  Also shown are the results from the H1 collaboration 
  (open squares) obtained from a measurement based on large rapidity gaps.}
\label{fg:eflow}
\end{figure}

Another measurement of the event topology is the distribution of energy as a
function of the pseudorapidity 
of a hadron with respect to the $\gamma^*\P$ CMS axis. 
This distribution, commonly referred to as the energy flow,
is shown in Fig.~\ref{fg:eflow} for data in three $M_X$ ranges.
The data for low $M_X$ have a Gaussian-like shape, but for $M_X > 7.5$~GeV a
structure with two peaks develops.
This structure becomes more pronounced as the mass increases. 
A slight asymmetry develops
in the data, with more energy being produced
in the $\gamma^*$ hemisphere.
For comparison, the 
predictions from the RAPGAP, SATRAP-CDM and RIDI Monte Carlo generators are also shown.
The SATRAP-CDM and RAPGAP generators predict too broad a rapidity distribution, 
and display a
separation between the $\gamma^*$ and \P~hemispheres in the lowest $M_X$ 
bin that is not exhibited by
the data. RIDI gives a reasonable description of the data in the lowest
$M_X$ bin but has a different shape at higher $M_X$.
The asymmetry indicated in the data is slightly larger
than that produced by the diffractive Monte Carlo models.
For those values of $|\eta|$ at which the H1 collaboration has also published
data~\cite{h1hadron}, 
the energy flow is in good agreement, except 
at the highest $M_X$ where the H1 data are somewhat narrower.

\begin{figure}[p]
\center{ \leavevmode \epsfxsize=17cm \epsffile{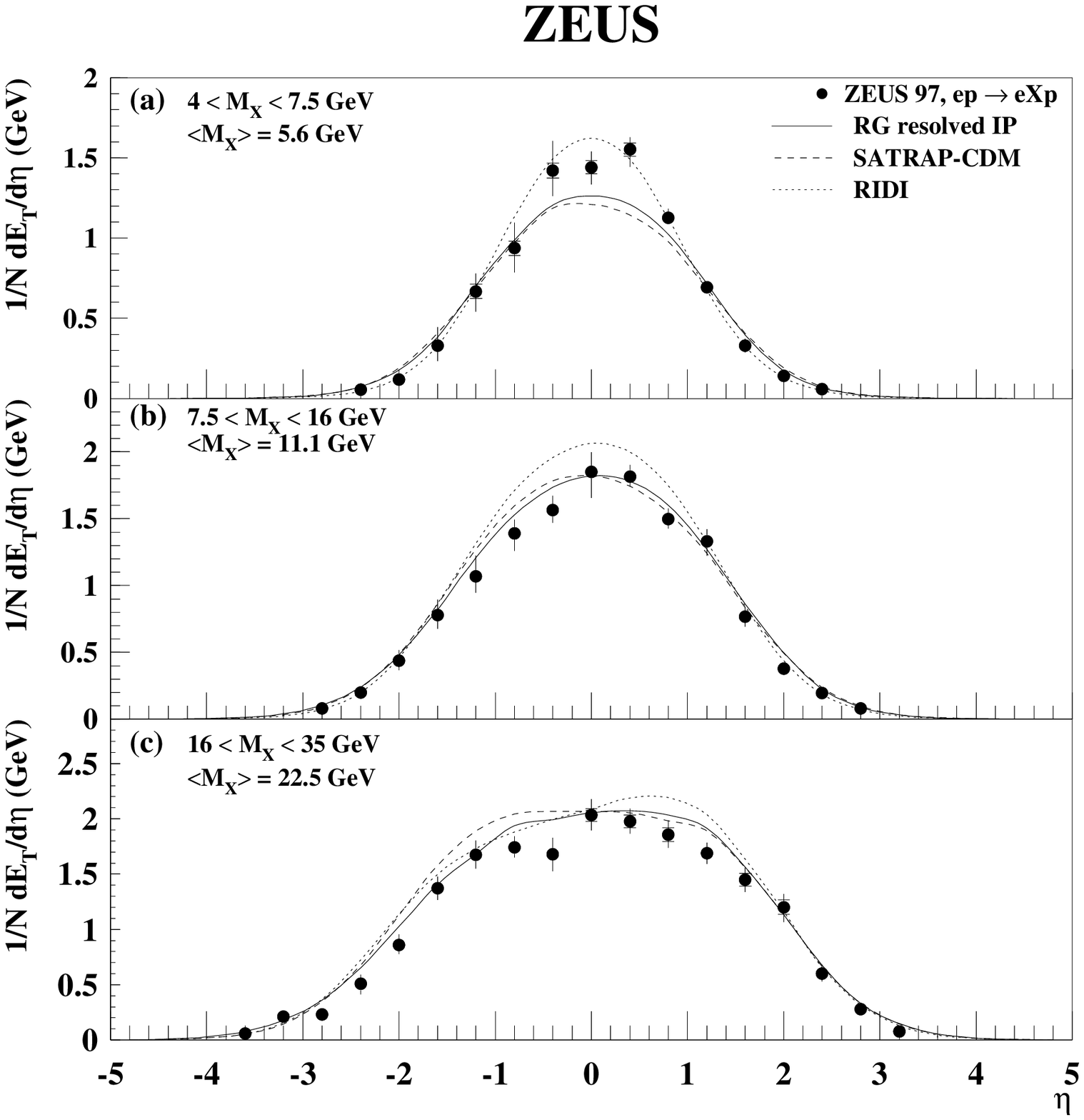} }
\center{ \leavevmode \epsfxsize=10cm \epsffile{PS_newfs/gam_pom_arrow2.eps} }
\caption[this space for rent]{
  The transverse energy flow 
  as a function of pseudorapidity, $\eta$, in the $\gamma^*\P$-CMS, for various
  ranges of $M_X$. The data are shown as solid points, while the Monte
  Carlo predictions from RAPGAP, SATRAP-CDM and RIDI are shown as solid, dashed and dotted curves,
        respectively.  Positive $\eta$ is in
  the direction of the $\gamma^*$.  The inner error bars show
  statistical uncertainties only; the outer bars show the statistical and
  systematic uncertainties added in quadrature.}
\label{fg:etflow}
\end{figure}

Figure \ref{fg:etflow} shows the distribution of transverse energy, 
$E_T = \sum E_i \sin{\theta_i} $, as a
function of $\eta$. Similar discrepancies between the data and the Monte Carlo 
events were observed for the
transverse-energy flow as was observed for the energy-flow distribution.


%
%

\section{Seagull distribution}
\label{sec:seagull}
 
The small asymmetry between positive and negative $\eta$ 
observed in the energy-flow plot of Fig.~\ref{fg:eflow}
can be further investigated using the distribution of transverse momentum
of the particles belonging to the system $X$.

In inclusive DIS, $ep \rightarrow eX$, the fragmentation of the hadronic system occurs
between the struck quark (forming the system X) and the remaining
quarks in the proton (forming the proton remnant). 
Measurements in the Breit frame \cite{disenergyflow} have shown that,
whereas the particle multiplicity and momentum distributions in the
hemisphere of the struck quark are roughly consistent with those
measured in $e^+ e^- \rightarrow q\bar{q}$, particles are
produced with smaller average transverse momentum in the proton-remnant hemisphere.

In diffractive DIS, an asymmetry in the momentum distribution  between the $\gamma^*$ 
and \P~hemispheres could be observed
if the Pomeron behaves as an extended object like the proton and produces a 
remnant after the scattering process.
This asymmetry is usually investigated by studying single-particle
distributions as a function of the scaled longitudinal momentum,
$x_F$, and the momentum transverse to the photon direction, $p_T$.
The photon direction defines the longitudinal axis in the
$\gamma^* \P$ CMS as well as in the $\gamma^* p$
CMS. If $p_T$ and $p_L$ are the momentum components
of a final-state hadron perpendicular and parallel, respectively, to this axis,
the variable $x_F$ is given by:
\begin{eqnarray}
x_F = p_L / {p_L^{max}} ,
\nonumber
\end{eqnarray}
where positive $x_F$ is in the direction of the $\gamma^*$ and
$p_L^{max}$ is the maximum kinematically allowed value of
$p_L$. In the $\gamma^* \P$ CMS, $p_L^{max} = M_X/2$.
For the $\gamma^* p$ case, $p_L^{max} = W/2$.

\begin{figure}[t]
\center{ \leavevmode \epsfxsize=15cm \epsffile{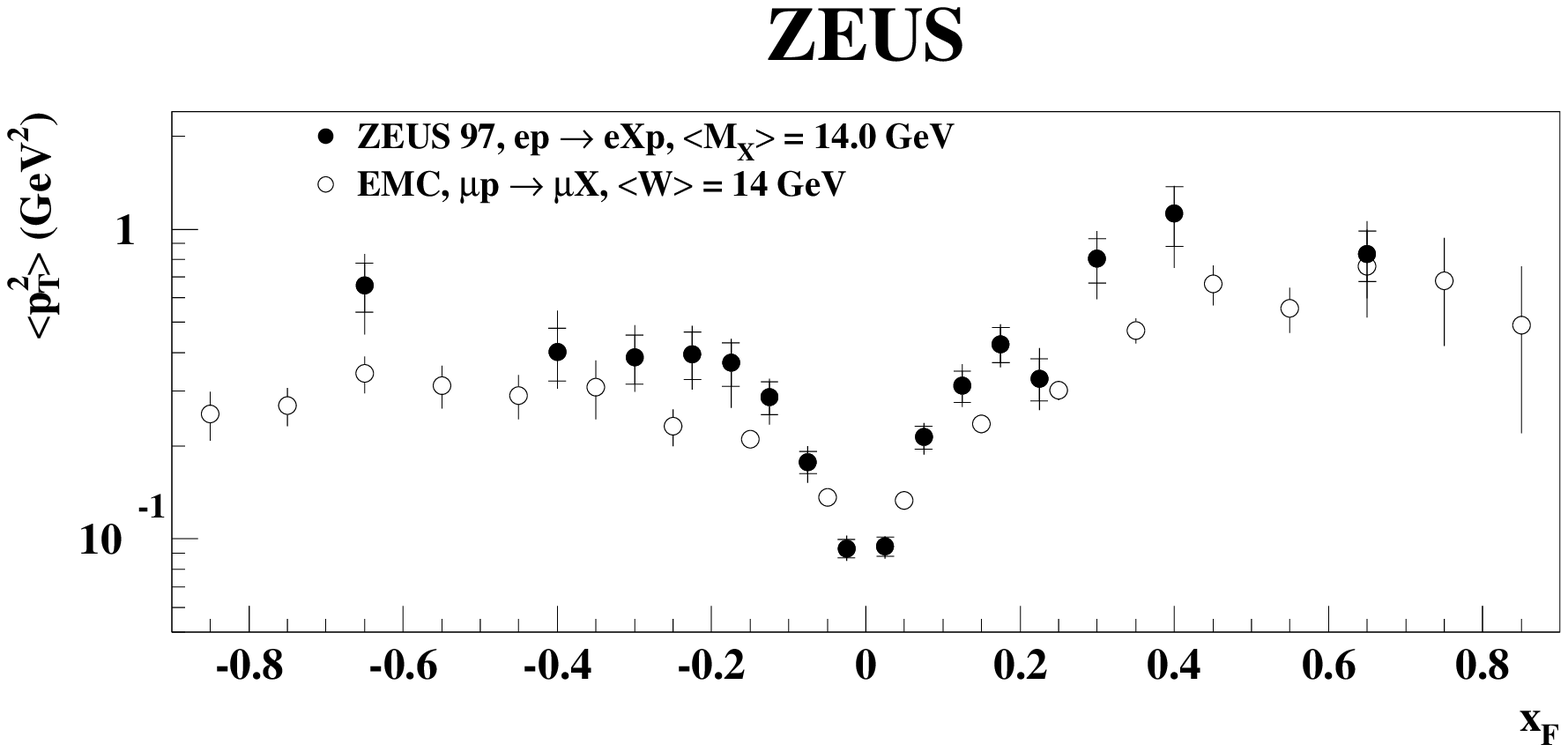} }
\center{ \leavevmode \epsfxsize=12cm \epsffile{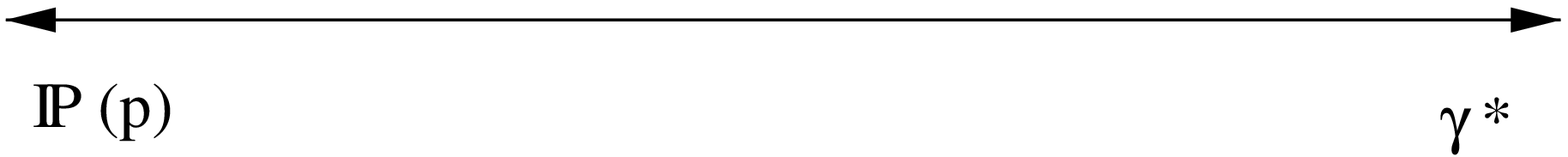} }
\caption[this space for rent]{
  Average squared transverse momentum of particles measured in the 
  center-of-mass frame of the system $X$ as a function of $x_F$
  for diffractive events from this analysis (solid circles)
  with $11 < M_X < 17.8$~GeV (\avM $ = 14.0$ GeV). 
Also shown (open circles)
  is the same quantity for inclusive DIS $\mu p \rightarrow \mu X$
  data from the EMC collaboration \cite{emc} at $W = $ \avM. 
Positive $x_F$ is in the direction of the  virtual photon. 
}
\label{fg:seagullemc}
\end{figure}

The average $p_T^2$ of particles
as a function of $x_F$, commonly referred to as the ``seagull plot'', 
is shown in Fig.~\ref{fg:seagullemc}
for $11 < M_X < 17.8$~GeV. Also plotted are $\gamma^* p$
data from the EMC \cite{emc} collaboration at $W = 14$~GeV, 
equal to the average value of $M_X$ in this bin. 
The EMC DIS $\mu p \rightarrow \mu X$ data indicate a
  suppression of the average $p_T^2$ associated with a proton
  remnant which is not as apparent in the diffractive data in this mass range.



\begin{figure}[p]
\begin{center}

\parbox[c]{18cm}{
        \epsfig{file=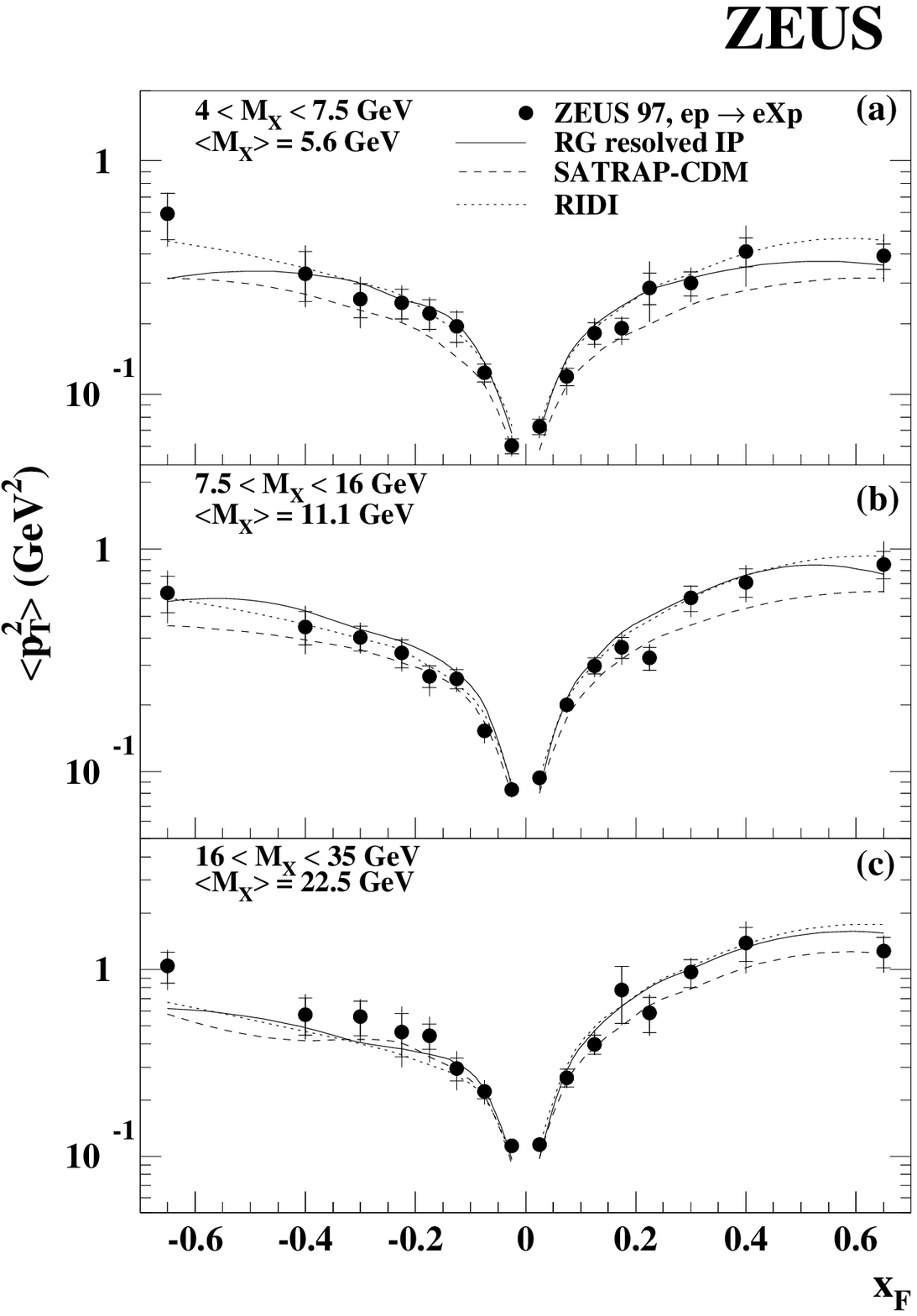, height=14.cm}
        \hspace{-1.5cm}
        \epsfig{file=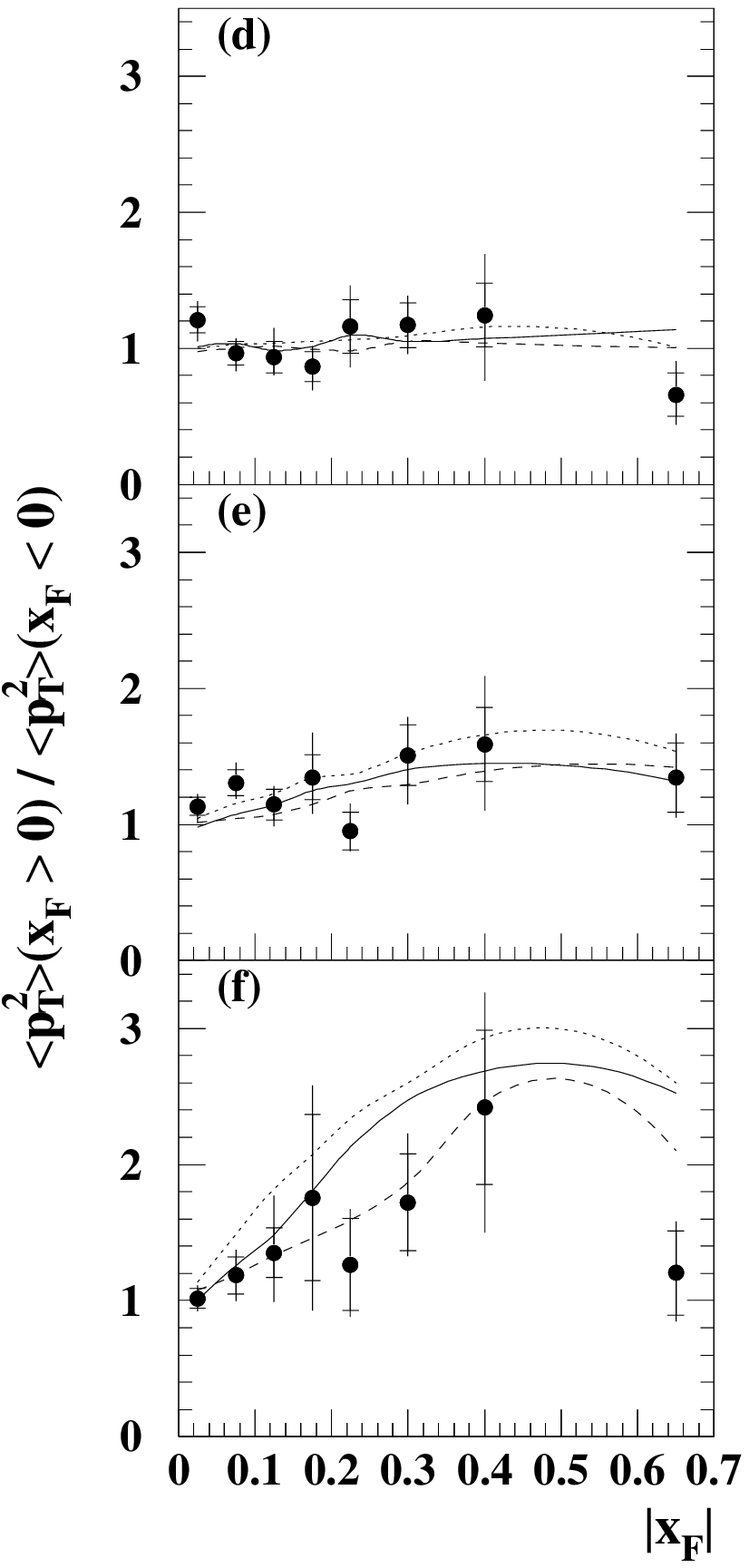, height=14.cm, width=8.cm}

        \hspace{+2.8cm}\epsfig{file=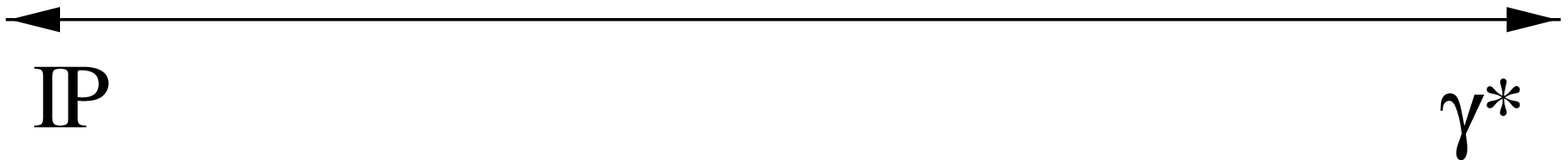, width=5.cm}
}

\caption[this space for rent]{
  Average squared transverse momentum as a function of $x_F$ (seagull plot)
  in three bins of $M_X$ in the $\gamma^* \P$-CMS (a-c) and the ratios of
  average momenta in the $\gamma^*$ and $\P$ hemisphere as function of $|x_F|$ (d-f). 
  The data (points) are compared to
  three models: RAPGAP (solid curve), SATRAP-CDM (dashed curve), and RIDI (dotted curve). 
  Positive $x_F$
  is in the direction of the virtual photon.
  The inner error bars show statistical uncertainties
  only; the outer bars show the statistical and systematic uncertainties
  added in quadrature.
}
\label{fg:seagull}
\end{center}
\end{figure}

Shown in Fig.~\ref{fg:seagull} (a)-(c) is the seagull plot
 for three different $M_X$ bins compared with the predictions from
RAPGAP, SATRAP-CDM and RIDI. 
The data exhibit a growing 
asymmetry as $M_X$ increases.
This asymmetry can also be seen in the ratios of the
average squared transverse-momentum in the $\gamma^*$ and \P~hemispheres
as a function of $|x_F|$ (Fig.~\ref{fg:seagull} (d)-(f)).
The data are well reproduced by both RAPGAP and RIDI, 
while for $M_X > 16$~GeV, the transverse momentum generated by SATRAP-CDM is
smaller than in the data.
RAPGAP and SATRAP-CDM describe the size of the
asymmetry correctly in all $M_X$ bins, while for $M_X > 16$~GeV RIDI 
slightly underestimates the transverse 
momentum in the \P~direction, resulting in a slightly larger asymmetry than that seen in the data.
RAPGAP produces this asymmetry by including a Pomeron remnant. 
RIDI and SATRAP-CDM, in contrast, produce the asymmetry via the $q \bar{q} g$ diagram.

%
%

\section{Conclusion}
\label{sec:conclude}

A study of the hadronic system, $X$,
 in the reaction $ep \rightarrow eXp$ has been reported for the  
kinematic range 
$4 < M_X < 35$~GeV, $4 < Q^2 < 150$~GeV$^2$, $0.0003 < \xp < 0.03$ and
$70 < W < 250$~GeV.
The use of the LPS allows diffractive events to be tagged without applying cuts
on the system $X$. It also provides a powerful constraint
on the diffractive kinematics, allowing, for example, an accurate
determination of the $\gamma^* \P$ axis in the 
center-of-mass frame of the system $X$.

The diffractive hadronic final state becomes more collimated as the invariant mass $M_X$ of the
system increases. This trend is similar to the one observed in $e^+e^-$ 
annihilation.
However, on average the diffractive final state is more isotropic.
This can be attributed to contributions 
not present in $e^+e^-$ annihilation, such as the boson-gluon fusion 
process in the resolved Pomeron approach, 
or $\qbarqg$ production from the dissociation of the virtual
photon.

The mean transverse momentum out of the event 
plane is similar to that found in
$e^+e^-$ annihilation, indicating the universality of the hadronization.
Even after considering the broadening effects of hadronization, it is
apparent that more $k_T$ than is usually associated with the
resolved Pomeron Monte Carlo approach is required to accommodate the large
thrust angle and narrow energy flows at low $M_X$.

Particle production becomes asymmetric along the $\gamma^*\P$~axis as
$M_X$ increases, resulting in more average transverse momentum in the
virtual-photon hemisphere. This asymmetry is consistent with both the concept 
of a remnant in the resolved Pomeron
model and with the production of $q\bar{q}g$ final states in the 
photon-dissociation approach.
The comparison with the Monte Carlo models suggests that a dominant gluon
contribution to the partonic final state is necessary.

The invariant mass, $M_X$, of the hadronic system is the only
variable upon which the characteristics of the system was found to depend. 
The system is independent of the
momentum transfer at either 
the positron vertex, $Q^2$, or at the proton vertex, $t$.
Neither is there any dependence on $x$ or on the fractional momentum of the 
Pomeron, $\xp$.

Many models of diffraction are able to reproduce the measured diffractive cross sections.
However, none of the models discussed here is able to describe
all aspects of the data. It is clear, therefore, that 
measurements of the detailed characteristics of 
diffractive events, such as presented in this paper, will become more and more 
crucial in understanding the underlying physics of diffraction in
deep inelastic scattering. 

%
%
\setcounter{secnumdepth}{0}
\section{Acknowledgments}
 
This measurement was made possible by the inventiveness and the diligent
efforts of the HERA machine group.
The strong support and encouragement of the DESY directorate has been
invaluable.
The design, construction, and installation of the ZEUS detector has been made
possible by the ingenuity and dedicated effort of many people from inside DESY
and from the home institutes who are not listed as authors. Their
contributions are acknowledged with great appreciation.
We acknowledge support by the following:
the Natural Sciences and Engineering Research Council of Canada (NSERC);
the German Federal Ministry for Education and
Science, Research and Technology (BMBF), under contract
numbers 057BN19P, 057FR19P, 057HH19P, 057HH29P, 057SI75I;
the MINERVA Gesellschaft f\"ur Forschung GmbH, the
Israel Science Foundation, the U.S.-Israel Binational Science
Foundation, the Israel Ministry of Science and the Benozyio Center
for High Energy Physics;
the German-Israeli Foundation, the Israel Science
Foundation, and the Israel Ministry of Science;
the Italian National Institute for Nuclear Physics (INFN);
the Japanese Ministry of Education, Science and
Culture (the Monbusho) and its grants for Scientific Research;
the Korean Ministry of Education and Korea Science
and Engineering Foundation;
the Netherlands Foundation for Research on Matter (FOM);
the Polish State Committee for Scientific Research,
grant no. 2P03B04616, 620/E-77/SPUB-M/DESY/P-03/DZ 247/2000 and
112/E-356/SPUB-M/DESY/P-03/DZ 3001/2000,
and by the German Federal Ministry for
Education and Science, Research and Technology (BMBF);
the Fund for Fundamental Research of Russian
Ministry for Science and Edu\-cation and by the German Federal Ministry for
Education and Science, Research and Technology (BMBF);
the Spanish Ministry of Education
and Science through funds provided by CICYT;
the Particle Physics and Astronomy Research Council, UK;
the US Department of Energy;
the US National Science Foundation.

%
%

\clearpage


\end{document}